\documentclass[useAMS,usenatbib,twocolumn]{mnras}

\usepackage{graphicx}
\usepackage{amssymb}
\usepackage{natbib}
\usepackage{cite}
\usepackage{textcomp}
\usepackage{pdflscape}

\bibpunct{(}{)}{;}{a}{}{,}
\usepackage{hyperref}

\hypersetup{draft}
\begin{document}

\title[On the origin of X-ray oxygen emission-lines in obscured AGN.]{On the origin of X-ray oxygen emission-lines in obscured AGN\thanks{Based on observations obtained with XMM-Newton, an ESA science mission with instruments and contributions directly funded by ESA Member States and NASA.}}
\author[V. Reynaldi et al.]{Reynaldi V.$^{1}$\thanks{E-mail:vreynaldi@fcaglp.unlp.edu.ar}, Guainazzi M.$^{2}$, Bianchi, S.$^{3}$, Andruchow, I.$^{1,4}$, Garc\'{\i}a, F.$^{5}$,
\newauthor Salerno, N.$^{1}$, L\'opez, I.E.$^{1}$ \\
$^{1}$Facultad de Ciencias Astron\'{o}micas y Geof\'{\i}sicas, UNLP, Paseo del Bosque s/n, La Plata 1900, Argentina\\
$^{2}$ESTEC/ESA, Keplerlaan 1, NL-2201AZ Noordwijk, the Netherlands\\
$^{3}$Dipartimento di Matematica e Fisica, Universit\`a degli Studi Roma Tre, via della Vasca Navale 84, I-00146 Roma, Italy\\
$^{4}$Instituto de Astrof\'{\i}sica de La Plata, CONICET, Paseo del Bosque s/n, La Plata 1900, Argentina\\
$^{5}$Kapteyn Astronomical Institute, University of Groningen, P.O. BOX 800, 9700 AV Groningen, The Netherlands
}

\date{}

\pagerange{\pageref{firstpage}--\pageref{lastpage}} \pubyear{}

\maketitle

\label{firstpage}

\begin{abstract}

We present the Catalog of High REsolution Spectra of Obscured Sources (CHRESOS) from the {\it XMM-Newton} Science Archive. It comprises the emission-line luminosities of H- and He-like transitions from C to Si, and the Fe 3C and Fe 3G L-shell ones. Here, we concentrate on the soft X-ray OVII~(f) and  OVIII~Ly$\alpha$ emission lines to shed light onto the physical processes with which their formation can be related to: active galactic nucleus vs. star forming regions. We compare their luminosity with that of two other important oxygen key lines [OIII]$\lambda5007$\AA, in the optical, and [OIV]25.89\micron, in the IR. We also test OVII~(f) and  OVIII~Ly$\alpha$ luminosities against that of continuum bands in the IR and hard X-rays, which point to different ionization processes. We probe into those processes by analyzing photoionization and colisional ionization model predictions upon our lines. We show that both scenarios can explain the formation and observed intensities of OVII~(f) and OVIII~Ly$\alpha$. By analyzing the relationships between OVII~(f) and OVIII~Ly$\alpha$, and all other observables: [OIII]$\lambda5007$\AA, [OIV]25.89\micron~emission lines, and MIR~12\micron, FIR~60\micron, FIR~100\micron, 2--10~keV and 14--195~keV continuum bands, we conclude that the AGN radiation field is mainly responsible of the soft X-ray oxygen excitation.

\end{abstract}


%%%%%%%%%%%%%%%%%%%%%%%%%%%%%%%%%%%%%%%%%%%%%%%%%%%%%%%%%%%%%%%%%%%%%%%%%%%%%%%%%%%
%
%           INTRODUCTION
%
%%%%%%%%%%%%%%%%%%%%%%%%%%%%%%%%%%%%%%%%%%%%%%%%%%%%%%%%%%%%%%%%%%%%%%%%%%%%%%%%%%%
\section{Introduction}
\label{intro}

X-ray spectroscopy of celestial sources can probe hot ({\it i.e.} $\sim$10$^{6}$~K) as well as cold ({\it i.e.} 10$^{3-4}$~K) plasma ionized by an external radiation field. However, only with the advent of grating and transmission spectrometers on board {\it Chandra} and {\it XMM-Newton}, X-ray spectroscopy has acquired a full maturity, allowing a complete diagnostic of the physical conditions and of the dynamics of plasma in a variety of astrophysical sources and contexts.

In this {\it paper} we primarily deal with X-ray spectroscopy of gas in the Narrow Line Regions (NLR) surrounding Active Galactic Nuclei (AGN). NLR gas is observationally charaterised in the optical band by emission lines of moderately-ionised species with a profile Full Width Half Maximum $\le$500~km~s$^{-1}$. The strongest observed spectral lines correspond to forbidden transitions such as [OIII]$\lambda$5007\AA\, and [NII]$\lambda$6584\AA\,. These lines are believed to be excited by the intense radiation field emitted by the nuclear accreting black hole \citep[][and references therein]{peterson97}. In nearby galaxies, NLRs are resolved to several hundreds or thousands of kpc (in the latter case called ``Extended Narrow Line Region'', ENLR).

The spectra of nearby obscured AGNs have systematically shown a strong soft X-ray excess above the extrapolation of the primary, absorbed nuclear continuum. High-resolution imaging and spectroscopy with {\it Chandra} and the Reflection Grating Spectrometer (RGS/{\it XMM-Newton}) have revealed that this component is dominated by strong recombination lines from He- and H-like transitions of light metals, as well as L transition of iron \citep{sako00_mrk3,kink02,guainazzi07}. The presence of narrow Radiative Recombination Continua, typical signatures of low temperature plasma \citep{liedahl96}, as well as plasma diagnostics based on the He-like triplets indicate that the X-ray emitting gas is photo-ionized, most likely by the AGN radiation field. Recently, Radiation Pressure Compression (RPC) has been proposed as a universal scenario for the gas in the AGN nuclear environment, from the sub-kpc scale of the Broad Line Regions and the torus \citep{stern14} to the ENLR \citep{bianchi19}. Deep high-resolution imaging with {\it Chandra} unveiled that a large fraction of the flux in the comparatively large  RGS aperture is extended on scales of a few arcseconds. This extended soft X-ray emission exhibit a striking morphological coincidence with the NLR as observed in the optical \citep{young01,bianchi06,levenson06}. While this evidence suggests a common origin between the optical and the X-ray emitting NLR gas, the ultimate proof of this connection is still to be found.

In this {\it paper} we report for the fist time on a systematic study of the correlation between spectroscopic measurements of He- and H-like oxygen transitions and multi-wavelength spectroscopic and photometric indicators of AGN or star-formation strength, aiming at understanding the ultimate origin of the X-ray ENLR. The results are based on the largest existing compilation of X-ray high-resolution spectra, the Catalogue of High REsolution Spectra of Obscured Sources (CHRESOS), which we present here. CHRESOS is an updated version of the sample first discussed in \citet{guainazzi07}. The {\it paper} is structured as follows: \S~\ref{data} presents CHRESOS and summarizes the employed archival data. In \S~\ref{sect:predictions} we compare the observed X-ray oxygen line ratios with prediction of state-of-the-art photoionized and optically-thin collisionally ionized models. \S~\ref{AGN} presents the results of our correlations. We discuss and summarize our main results in \S~\ref{conclusions}.

%%%%%%%%%%%%%%%%%%%%%%%%%%%%%%%%%%%%%%%%%%%%%%%%%%%%%%%%%%%%%%%%%%%%%%%%%%%%%%%%%%%
%
%           OBSERVATIONS
%
%%%%%%%%%%%%%%%%%%%%%%%%%%%%%%%%%%%%%%%%%%%%%%%%%%%%%%%%%%%%%%%%%%%%%%%%%%%%%%%%%%%

\section{Data.}
\label{data}

CHRESOS comprises 100 sources of the local Universe (redshift $z<0.07$) that harbor a Seyfert-type Active Galactic Nucleus (AGN) according to {\it XMM-Newton} observations in the soft X-ray band. It gathers spectra obtained by the Reflection Grating Spectrometer \citep{herder01} onboard {\it XMM-Newton}, sensitive in the 0.2--2.0~keV~energy band, available in the public archive. In this paper we present the Catalogue with the emission-line luminosities in the soft X-ray band. The details on reduction procedure, spectral analysis and emission-line measurements are fully described in \citet{bianchi19}, where part of the CHRESOS sample has been used in the context of RPC analysis. Briefly, each line was fit with a Gaussian profile on top of a power-law continuum. A given line is considered to be detected if it yields an improvement in the quality of the fit at the 3$\sigma$ confidence level for one interesting parameter \citep{lampton76}. CHRESOS includes H- and He-like transitions from C to Si, and two L-shell emission lines from FeXVII: 3G and 3C. The emission lines and their rest frame wavelenghts (\AA) and energies (keV) are listed in Table~\ref{line-list}. We present the observed emission-line luminosities (with their errors quoted at 1$\sigma$ level) for each source in two separate tables. The H-like and Fe lines are listed in Table~\ref{H-like}. The He-like transition forms a triplet composed by a resonant, an  intercombination, and a forbidden line; we present the three emission lines for the He-like oxygen triplet, as well as the forbidden lines for the remaining species. These He-like emission lines are listed in Table~\ref{He-like}.

%%%%%%%%%%%%%%%%%%%%%%%%%%%%%%%%%%%%%%%%%%%%%%%%%%%%%%%%%%%%%%%%%%%%%%%%%%%%%%%%%%%%%%%%%%%%%%%%%%%%%%%%
\begin{table}
\caption{List of emission lines within CHRESOS Catalogue.}\label{line-list}
\begin{tabular}{|l|c|c|}
\hline
Emission line & Wavelength ($\AA$) & Energy (keV) \\
\hline
C V He$\beta$ &     34.973 & 0.355 \\ 
N VI (f) &          29.534 & 0.420 \\
N VII Ly$\alpha$ &  24.780 & 0.5 \\
O VII (f)         & 22.101  &  0.561  \\
O VII (i)         & 21.806  &  0.569  \\
O VII (r)         & 21.602  &  0.574  \\
O VIII Ly$\alpha$ & 18.967  &  0.654  \\
Fe XVII 3G       &  17.054  &  0.727  \\
Fe XVII 3C       &  15.015  &  0.826  \\
Ne IX (f)         & 13.699  &  0.905  \\
Ne X Ly$\alpha$   & 12.132  &  1.02   \\
Mg XI (f)         & 9.23    &  1.343  \\
Mg XII Ly$\alpha$ & 8.419   &  1.473  \\
Si XIII (f)       & 6.69    & 1.853   \\
Si XIV Ly$\alpha$ & 6.181   & 2.006 \\
\hline
\end{tabular}
\end{table}

%%%%%%%%%%%%%%%%%%%%%%%%%%%%%%%%%%%%%%%%%%%%%%%%%%%%%%%%%%%%%%%%%%%%%%%%%%%%%%%%%%%%%%%%%%%%%%%%%%%%%%%

As long as this article is concerned, we are going to use oxygen emission lines: the forbidden component from the He-like OVII triplet [OVII(f)], and OVIII~Ly$\alpha$, the H-like transition. We will compare them with multi-wavelength nuclear data, obtained from the literature: continuum luminosities in 14--195~keV, 2--10~keV, Mid Infrared (MIR) 12\micron, Far Infrared (FIR) 60\micron, and FIR 100\micron, and luminosities of two other important oxygen lines: [OIII]$\lambda5007$\AA~ in the optical, and [OIV]25.89\micron~ in the IR. Since we aim to probe into the formation mechanism of OVII (f) and OVIII~Ly$\alpha$, those multi-wavelength data were chosen because of their relationships with the two scenarios from where these emission lines can emerge: the AGN and the (nuclear/near nuclear) star-forming regions, or starbursts (SB){\citep{guainazzi07}}. 

The 2--10~keV and the 14--195~keV (the {\it Swift}/BAT band) are the most reliable measurements of the nuclear (primary) continuum in non-Compton-thick sources \citep{weaver10}. The Table~\ref{He-like} also includes the 2--10~keV absorption-corrected luminosities and references from the literature. When the data was not available, we obtained the intrinsic luminosities as part of this work through the EPIC/MOS spectra available in the XMM-Newton Science Archive. The 14--195~keV data were obtained from the {\it Swift}/BAT 105-months Hard X-ray Survey \citep{oh18}.

%%%%%%%%%%%%%%%%%%%%%%%%%%%%%%%%%%%%%%%%%%%%%%%%%%%%%%%%%%%%%%%%%%%%%%%%%%%%%%%%%%%

\begin{landscape}
\begin{table}
\caption{CHRESOS Catalogue: H-like emission-lines and L-shell transitions of Fe.}\label{H-like}
\begin{tabular}{|l|c|c|c|c|c|c|c|c|c}
\hline
Source & CVI~Ly$\alpha$ & CV~He$\beta$ & NVII~Ly$\alpha$ & OVIII~Ly$\alpha$ & NeX~Ly$\alpha$ & MgXII~Ly$\alpha$ & SiXIV~Ly$\alpha$ & Fe 3C   & Fe 3G\\
{\sc i}&  {\sc ii}      &   {\sc iii}  &     {\sc iv}    &      {\sc v}     &    {\sc vi}    &    {\sc vii}     &     {\sc viii}   &{\sc ix} & {\sc x}\\
\hline
CIRCINUS & --- & --- & --- & --- & $37.95_{-0.100}^{+0.08}$ & $38.05_{-0.09}^{+0.08}$ & $38.45_{-0.16}^{+0.12}$ & $ 38.12^{+0.41}$ & --- \\ 
ESO138-G01 & --- & --- & --- & $39.49_{-0.100}^{+0.08}$ & --- & $39.45_{-0.17}^{+0.12}$ & --- & --- & --- \\
ESO323-G077 & --- & --- & --- & $39.52_{-0.14}^{+0.09}$ & --- & --- & --- & --- & --- \\
ESO362-G018 & $39.99_{-0.07}^{+0.05}$ & --- & $39.46_{-0.11}^{+0.09}$ & $39.85_{-0.05}^{+0.05}$ & --- & --- & --- & --- & --- \\
H0557-385 & --- & --- & --- & --- & --- & --- & $41.32_{-0.06}^{+0.06}$ & --- & --- \\
IRAS05189-2524 & --- & --- & $40.14_{-0.18}^{+0.14}$ & $40.51_{-0.12}^{+0.05}$ & --- & --- & --- & --- & --- \\
MCG-03-34-064 & --- & --- & --- & $39.80_{-0.20}^{+0.14}$ & --- & --- & --- & $39.83^{+0.37}$ & --- \\
MCG-03-58-07 & --- & --- & --- & --- & --- & --- & $41.08_{-0.12}^{+0.14}$ & --- & --- \\
MRK231 & --- & --- & $40.03_{-0.19}^{+0.21}$ & --- & $40.42_{-0.17}^{+0.14}$ & --- & --- & --- & --- \\
MRK268 & --- & --- & --- & --- & --- & --- & $41.92_{-0.16}^{+0.04}$ & --- & --- \\
MRK273 & --- & --- & --- & $40.28_{-0.21}^{+0.23}$ & --- & --- & --- & --- & --- \\
MRK3 & --- & --- & --- & $40.16_{-0.05}^{+0.04}$ & $39.85_{-0.13}^{+0.101}$ & --- & --- & $39.4^{+0.43}$ & --- \\
MRK477 & --- & --- & --- & $40.31_{-0.32}^{+0.19}$ & --- & --- & --- & --- & --- \\
NGC1052 & --- & --- & $38.33_{-0.16}^{+0.12}$ & $38.62_{-0.09}^{+0.08}$ & --- & --- & --- & --- & $38.33^{+0.36}$ \\
NGC1068 & $40.24_{-0.008}^{+0.011}$ & $39.52_{-0.04}^{+0.03}$ & $40.06_{-0.006}^{+0.014}$ & $40.13_{-0.008}^{+0.008}$ & $39.65_{-0.012}^{+0.03}$ & $39.28_{-0.07}^{+0.06}$ & --- & $39.65^{+0.06}$ & $39.74^{+0.06}$ \\
NGC1320 & --- & --- & $39.36_{-0.23}^{+0.16}$ & --- & --- & --- & --- & --- & --- \\
NGC1365 & $39.08_{-0.098}^{+0.02}$ & --- & $39.08_{-0.03}^{+0.02}$ & $39.32_{-0.02}^{+0.02}$ & $38.94_{-0.06}^{+0.06}$ & --- & --- & --- & --- \\
NGC2655 & --- & --- & --- & $39.06_{-0.14}^{+0.11}$ & --- & --- & --- & --- & --- \\
NGC3393 & --- & --- & --- & $39.77_{-0.19}^{+0.14}$ & --- & --- & --- & --- & --- \\
NGC4151 & $39.50_{-0.008}^{+0.009}$ & $38.81_{-0.06}^{+0.04}$ & $39.13_{-0.005}^{+0.02}$ & $39.54_{-0.005}^{+0.005}$ & $38.73_{-0.03}^{+0.03}$ & $38.52_{-0.09}^{+0.07}$ & $39.03_{-0.16}^{+0.11}$ & $38.44^{0.14}$ & $38.58^{+0.14}$ \\
NGC424 & --- & --- & --- & $39.30_{-0.12}^{+0.09}$ & $39.32_{-0.16}^{+0.12}$ & --- & --- & --- & --- \\
NGC4388 & --- & --- & --- & $39.46_{-0.15}^{+0.12}$ & --- & --- & --- & --- & --- \\
NGC4507 & $39.80_{-0.17}^{+0.13}$ & --- & --- & $39.81_{-0.07}^{+0.06}$ & --- & --- & --- & --- & $39.31^{+0.30}$\\
NGC5252 & --- & --- & --- & $39.75_{-0.21}^{+0.15}$ & --- & --- & --- & --- & --- \\
NGC5506 & --- & --- & $38.44_{-0.14}^{+0.11}$ & $39.15_{-0.04}^{+0.04}$ & $38.79_{-0.13}^{+0.10}$ & --- & --- & $38.81^{+0.17}$ & $38.71^{+0.21}$ \\
NGC5548 & $39.87_{-0.10}^{+0.08}$ & --- & --- & $40.02_{-0.04}^{+0.04}$ & --- & --- & --- & --- & --- \\
NGC5643 & --- & --- & --- & $38.38_{-0.16}^{+0.12}$ & --- & --- & --- & --- & --- \\
NGC6240 & --- & --- & --- & $40.44_{-0.1}^{+0.08}$ & $40.37_{-0.16}^{+0.17}$ & --- & --- & --- & --- \\
NGC7172 & --- & --- & --- & $38.92_{-0.20}^{+0.07}$ & --- & --- & --- & --- & --- \\
NGC7582 & --- & --- & $38.62_{-0.1}^{+0.08}$ & $38.98_{-0.04}^{+0.04}$ & $38.75_{-0.11}^{+0.09}$ & --- & --- & $38.86^{+0.21}$ & $38.63^{+0.31}$\\
NGC777 & --- & --- & --- & $40.12_{-0.10}^{+0.10}$ & $40.16_{-0.15}^{+0.174}$ & --- & --- & --- & --- \\
PG1411+442 & --- & --- & --- & --- & $41.37_{-0.14}^{+0.22}$ & --- & --- & --- & --- \\
PG1535+547 & --- & --- & --- & $40.05_{-0.13}^{+0.09}$ & --- & --- & --- & --- & --- \\
UGC1214 & --- & --- & --- & $40.11_{-0.18}^{+0.14}$ & --- & --- & --- & --- & --- \\
\hline
\end{tabular}
\flushleft {\it Notes}: ({\sc i}) Name of the source. ({\sc ii}) to ({\sc x}) Line luminosities in units of log(erg$\cdot$ s$^{-1}$; the errors on the line luminosity are quoted at the 1$\sigma$ level \citep{bianchi19}. Places marked as ``--" represent undetected lines.
\end{table}
\end{landscape}

%%%%%%%%%%%%%%%%%%%%%%%%%%%%%%%%%%%%%%%%%%%%%%%%%%%%%%%%%%%%%%%%%%%%%%%%%%%%%%%%%%%%%%%%%%%%%%%%%%%%%%%%%%%%%%%%%%%%%%%%%%%%%%%%%%%%%%%%%%%
\begin{landscape}
\begin{table}
\centering
\caption{CHRESOS Catalogue: He-like oxygen triplet; the forbidden emission-line of He-like triplets is the only one shown for the remaining species.}\label{He-like}
\begin{tabular}{|l|c|c|c|c|c|c|c|c|c}
\hline
Source & NVI~(f) &  OVII~(f) & OVII~(i) & OVII~(r) & NeIX~(f) & MgXI~(f) & SiXIII~(f) & 2--10~keV & Ref.\\
{\sc i}&{\sc ii} &{\sc iii}  &{\sc iv}  &  {\sc v} & {\sc vi} & {\sc vii} & {\sc viii}& {\sc ix}  & {\sc x}\\
\hline
2MASXJ12384342+09273 & --- & --- & --- & --- & $41.58_{-0.2}^{+0.11}$ & --- & --- & 43.5 & 1 \\
CIRCINUS & --- & --- & --- & --- & $37.78_{-0.15}^{+0.12}$ & $37.60_{-0.22}^{+0.15}$ & $37.76_{-1.19}^{+0.12}$ & 42.63 & 2 \\
ESO103-G35 & --- & $39.85_{-0.32}^{+0.26}$ & --- & --- & --- & --- & --- & 43.37 & 2 \\
ESO104-G11 & --- & --- & --- & --- & $39.24_{-1.16}^{+0.30}$ & --- & --- & 40.7 & 1 \\
ESO137-G34 & --- & --- & --- & --- & $38.73_{-1.16}^{+0.64}$ & --- & --- & 42.64 & 2 \\
ESO138-G01 & --- & $39.70_{-0.12}^{+0.1}$ & $39.31_{-0.32}^{+0.37}$ & $39.10_{-0.43}^{+0.68}$ & $39.32_{-0.10}^{+0.08}$ & --- & --- & 44.09 & 2 \\
ESO323-G077 & --- & --- & --- & --- & $39.08_{-0.48}^{+0.19}$ & --- & --- & 42.87 & 2 \\
ESO362-G018 & $39.48_{-0.14}^{+0.11}$ & $40.18_{-0.04}^{+0.04}$ & $36.32_{-0.43}^{+589.71}$ & $39.65_{-0.40}^{+0.32}$ & $39.47_{-0.12}^{+0.09}$ & --- & --- & 42.96 & 2 \\
ESO383-G18 & --- & $39.90_{-0.22}^{+0.16}$ & --- & --- & $39.17_{-695.0}^{+0.26}$ & --- & --- & 42.78 & 2 \\
F00521-7054 & --- & --- & --- & --- & --- & $41.16_{-0.90}^{+0.29}$ & --- & 43.43 & 3 \\
FBQSJ075800.0+392029 & --- & $41.66_{-0.13}^{+0.13}$ & --- & --- & --- & --- & --- & 43.83 & 1 \\
H0557-385 & --- & $40.59_{-0.16}^{+0.10}$ & --- & --- & --- & $40.05_{-0.82}^{+0.50}$ & --- & 44.08 & 2 \\
IC1867 & --- & --- & --- & --- & $40.13_{-0.57}^{+0.24}$ & --- & --- & 41.21 & 1 \\
IGRJ19473+4452 & --- & --- & --- & --- & --- & $40.92_{-0.56}^{+0.13}$ & --- & 43.93 & 1 \\
IRAS01475-0740 & --- & --- & --- & --- & $39.52_{-0.77}^{+0.23}$ & --- & --- & 42.04 & 4 \\
IRAS04507+0358 & --- & --- & --- & --- & $40.03_{-1.4}^{+0.25}$ & $40.43_{-0.71}^{+0.18}$ & --- & 44.0 & 2 \\
IRAS05189-2524 & --- & $40.18_{-0.39}^{+0.22}$ & --- & --- & $40.18_{-0.21}^{+0.12}$ & --- & $40.44_{-0.49}^{+0.20}$ & 43.40 & 2 \\
IRAS15480-0344 & --- & --- & --- & --- & --- & $39.93_{-0.88}^{+0.25}$ & --- & 43.0 & 5 \\
IRAS18325-5926 & --- & $39.49_{-0.36}^{+0.20}$ & --- & --- & --- & --- & --- & 43.3 & 6 \\ 
MCG-01-05-047 & --- & --- & --- & --- & $39.30_{-0.33}^{+0.60}$ & --- & --- & 42.74 & 2 \\
MCG-03-34-064 & --- & $40.34_{-0.11}^{+0.09}$ & $39.62_{-0.33}^{+0.37}$ & $39.61_{-0.29}^{+0.32}$ & --- & --- & --- & 43.4 & 2 \\ 
MCG-03-58-07 & --- & --- & --- & --- & $39.70_{-0.46}^{+0.16}$ & --- & --- & 42.78 & 5 \\
MRK 231 & --- & --- & --- & --- & --- & --- & --- & 42.59 & 2 \\
MRK 268 & --- & --- & --- & --- & --- & --- & --- & 43.55 & 2 \\
MRK 273 & --- & --- & --- & --- & --- & --- & --- & 42.08 & 4 \\
MRK417 & --- & --- & --- & --- & --- & $40.85_{-0.30}^{+0.19}$ & --- & 43.72 & 2 \\
MRK1298 & --- & $40.9_{-0.17}^{+0.15}$ & $39.93_{-0.43}^{+1.64}$ & $40.38_{-0.43}^{+0.33}$ & --- & --- & --- & 43.3 & 7 \\ 
MRK3 & $39.35_{-0.68}^{+0.28}$ & $40.18_{-0.07}^{+0.06}$ & $39.46_{-0.23}^{+0.25}$ & $39.95_{-0.08}^{+0.11}$ & $39.75_{-0.10}^{+0.08}$ & --- & --- & 43.67 & 2 \\
MRK463 & $40.67_{-0.50}^{+0.26}$ & --- & --- & --- & $40.54_{-0.27}^{+0.24}$ & --- & --- & 43.09 & 2 \\
MRK477 & --- & $40.85_{-0.19}^{+0.15}$ & --- & --- & $40.14_{-0.57}^{+0.24}$ & --- & --- & 43.26 & 2 \\
MRK6 & --- & $39.69_{-0.50}^{+0.23}$ & --- & --- & --- & --- & --- & 43.08 & 2 \\
MRK704 & $40.66_{-0.10}^{+0.10}$ & $40.96_{-0.11}^{+0.09}$ & $40.45_{-0.18}^{+0.19}$ & $39.45_{-0.43}^{+1.68}$ & $40.33_{-0.19}^{+0.10}$ & --- & --- & 43.33 & 2 \\ 
NGC1052 & --- & $38.33_{-0.22}^{+0.16}$ & --- & --- & $38.07_{-0.32}^{+0.19}$ & --- & --- & 41.62 & 2 \\
NGC1068 & $40.16_{-0.005}^{+0.02}$ & $40.37_{-0.009}^{+0.01}$ & $39.70_{-0.03}^{+0.03}$ & $40.18_{-0.02}^{+0.02}$ & $39.65_{-0.02}^{+0.03}$ & $39.07_{-0.09}^{+0.07}$ & $39.10_{-0.28}^{+0.10}$ & 42.93 & 2 \\ 
NGC1320 & --- & $39.71_{-0.17}^{+0.14}$ & --- & --- & $39.20_{-0.28}^{+0.18}$ & --- & --- & 42.85 & 2 \\
NGC1365 & --- & $39.16_{-0.02}^{+0.05}$ & --- & --- & $38.90_{-0.06}^{+0.04}$ & $38.32_{-0.48}^{+0.22}$ & --- & 42.32 & 2 \\
NGC2110 & --- & --- & --- & --- & $38.58_{-1.55}^{+0.31}$ & --- & --- & 42.68 & 2 \\
NGC2655 & --- & --- & --- & --- & --- & --- & --- & 41.34 & 2 \\
NGC3227 & $37.86_{-0.83}^{+0.27}$ & $38.53_{-0.12}^{+0.09}$ & $38.40_{-0.23}^{+0.25}$ & $36.84_{-0.43}^{+5.79}$ & --- & --- & --- & 42.07 & 2 \\
NGC3393 & --- & $40.08_{-0.20}^{+0.12}$ & --- & --- & --- & $39.75_{-0.33}^{+0.19}$ & --- & 42.63 & 2 \\
NGC4151 & $39.30_{-0.009}^{+0.010}$ & $39.93_{-0.004}^{+0.006}$ & $39.26_{-0.02}^{+0.02}$ & $39.54_{-0.014}^{+0.014}$ & $39.15_{-0.012}^{+0.012}$ & $38.57_{-0.06}^{+0.05}$ & --- & 42.31 & 2 \\
NGC424 & $39.49_{-0.17}^{+0.06}$ & $40.0_{-0.05}^{+0.04}$ & --- & --- & $39.07_{-0.18}^{+0.13}$ & --- & --- & 43.77 & 2 \\

\hline
\end{tabular}
\end{table}
\end{landscape}

\begin{landscape}
\begin{table}
\centering
\contcaption{}%\label{He-like}
\begin{tabular}{|l|c|c|c|c|c|c|c|c|c}
\hline
Source & NVI~(f) &  OVII~(f) & OVII~(i) & OVII~(r) & NeIX~(f) & MgXI~(f) & SiXIII~(f)& 2--10~keV & Ref.\\
{\sc i}&{\sc ii} &{\sc iii}  &{\sc iv}  &  {\sc v} & {\sc vi} & {\sc vii} & {\sc viii}& {\sc ix}  & {\sc x}\\\\
\hline
NGC4388 & --- & $39.70_{-0.13}^{+0.11}$ & --- & --- & $38.88_{-3.00}^{+0.32}$ & --- & --- & 43.05 & 2 \\
NGC4395 & --- & $36.98_{-0.24}^{+0.16}$ & --- & --- & --- & --- & --- & 40.45 & 2 \\
NGC4507 & $39.23_{-0.41}^{+0.24}$ & $40.34_{-0.05}^{+0.04}$ & $39.29_{-0.24}^{+0.27}$ & $39.87_{-0.08}^{+0.09}$ & $39.59_{-0.09}^{+0.08}$ & $39.05_{-0.82}^{+0.28}$ & --- & 43.51 & 2 \\ 
NGC513 & --- & --- & --- & --- & --- & --- & $40.62_{-0.27}^{+0.14}$ & 42.66 & 2 \\
NGC5252 & --- & $40.34_{-0.15}^{+0.09}$ & $39.21_{-0.43}^{+0.97}$ & $40.29_{-0.14}^{+0.08}$ & $39.81_{-0.26}^{+0.11}$ & --- & --- & 43.01 & 2 \\
NGC5273 & $38.07_{-0.64}^{+0.28}$ & $38.65_{-0.18}^{+0.14}$ & $38.52_{-0.43}^{+0.26}$ & $38.46_{-0.43}^{+0.26}$ & --- & --- & --- & 41.25 & 2 \\ 
NGC5506 & --- & $38.88_{-0.102}^{+0.08}$ & $38.41_{-0.23}^{+0.26}$ & $38.09_{-0.43}^{+0.57}$ & $38.61_{-0.11}^{+0.09}$ & --- & --- & 42.99 & 2 \\ 
NGC5548 & $39.74_{-0.09}^{+0.08}$ & $40.59_{-0.02}^{+0.02}$ & --- & --- & $39.69_{-0.09}^{+0.07}$ & $39.32_{-1.16}^{+0.28}$ & $39.93_{-0.96}^{+0.18}$ & 43.14 & 2 \\
NGC5643 & $37.99_{-0.92}^{+0.29}$ & $38.70_{-0.13}^{+0.11}$ & $35.94_{-0.43}^{+45.84}$ & $38.65_{-0.20}^{+0.23}$ & --- & $37.97_{-0.83}^{+0.29}$ & --- & 42.43 & 2 \\
NGC6240 & --- & --- & --- & --- & $39.81_{-0.52}^{+0.24}$ & --- & --- & 44.75 & 2 \\
NGC7172 & --- & --- & --- & --- & --- & --- & --- & 42.74 & 2 \\
NGC7582 & --- & $38.78_{-0.11}^{+0.07}$ & $38.72_{-0.16}^{+0.17}$ & $38.92_{-0.10}^{+0.10}$ & $38.33_{-0.17}^{+0.13}$ & $38.51_{-0.25}^{+0.16}$ & --- & 43.48 & 2 \\
NGC777 & --- & --- & --- & --- & $40.84_{-0.03}^{+0.02}$ & --- & --- & 40.97 & 1 \\
NVSS193013+341047 & --- & --- & --- & --- & $40.74_{-1.55}^{+0.32}$ & --- & --- & 43.95 & 2\\
PG1411+442 & --- & --- & --- & --- & --- & $41.24_{-0.42}^{+0.17}$ & --- & 43.41 & 8 \\
PG1535+547 & --- & --- & --- & --- & $39.74_{-0.38}^{+0.22}$ & --- & --- & 42.06 & 9 \\
UGC05101 & --- & --- & --- & --- & --- & $40.25_{-1.05}^{+0.19}$ & --- & 44.2 & 2 \\
UGC1214 & --- & $40.45_{-0.14}^{+0.12}$ & --- & --- & $39.51_{-0.71}^{+0.30}$ & --- & --- & 41.3 & 10 \\
UGC3142 & --- & --- & --- & --- & --- & --- & $40.74_{-0.72}^{+0.16}$ & 43.05 & 11 \\
\hline
\end{tabular}
\flushleft {\it Notes}: ({\sc i}) Name of the source. ({\sc ii}) to ({\sc viii}) Line luminosities in units of log(erg$\cdot$ s$^{-1}$); the errors on the line luminosity are quoted at the 1$\sigma$ level \citep{bianchi19}. ({\sc ix})-({\sc x}) Unabsorbed intrinsic 2--10keV luminosities (same units) and their references as follows: [1]: this work; [2] \citet{ricci17}; [3]: \citet{tan12}; [4]: \citet{huang11}; [5]: \citet{brightman11}; [6]: \citet{iwasawa04}; [7]: \citet{giustini11}; [8]: \citet{zhou10}; [9]: \citet{inoue07}; [10]: \citet{bianchi10}; [11]: \citet{ricci10}. Places marked as ``--" represent undetected lines (see text).
\end{table}
\end{landscape}

%%%%%%%%%%%%%%%%%%%%%%%%%%%%%%%%%%%%%%%%%%%%%%%%%%%%%%%%%%%%%%%%%%%%%%%%%%%%%%%%%%%%%%%%%%%%%%%%%%%%%%%%%%%%%%%%%%%%%%%

%...... [OIII]
The optical [OIII]$\lambda5007$ emission line is an indicator of the AGN ionizing power. It forms in the NLR, the hundred-parsec scale nebular structure whose distribution, when mapped by [OIII]$\lambda5007$\AA, usually coincides with the gas distribution in the soft X-ray band \citep{young01,bianchi06,bianchi10,levenson06,dadina10,fabbiano18}. Due to its extention, the NLR emission do not suffer significant dust obscuration from the toroidal structure that surrounds the innermost regions of the AGN \citep{lamassa10}. The use of [OIII]$\lambda5007$ as a proxy of the AGN is supported by the very studied relationship between [OIII]$\lambda5007$ and the 2--10~keV luminosities \citep{mulchaey94,bassani99}. The precise form of such a relation is strongly dependent on whether the data are selected from the optical or from X-rays \citep{heckman05}; and it has also been argued that, in some particular cases, the relationship is even absent \citep{berney15,rojas17}. However, in the vast majority of cases, for all kind of AGNs (from Seyferts to 3C sources), the luminosity of [OIII]$\lambda5007$ emission line shows a strong and tight relationship with the hard X-ray 2--10~keV band, and is therefore widely accepted as an isotropic indicator of the primary continuum luminosity \citep[e.g.][]{mulchaey94,bassani99,schmitt03,heckman05,melendez08,hardcastle09,lamassa10,zhang17}.

%....... [OIV]
The [OIV]25.89\micron~IR emission line forms in the NLR too \citep{lamassa10}. The emission of [OIV]25.89\micron~and [OIII]$\lambda5007$ from nearby ($z<0.08$) Seyfert sources was correlated against the two hard X-ray nuclear continuum bands: 2--10~keV and 14--195~keV by \citet{melendez08}. Although they noted that the relationship that arises from [OIV]25.89\micron~vs~2--10~keV is weaker than the well-known [OIII]$\lambda5007$~vs~2--10~keV relationship, they also showed that the two oxygen emission lines are well correlated with the BAT band. They concluded that [OIV]25.89\micron~will be an isotropic proxy of the AGN power as long as the star formation can be neglected. In the same sense, \citet{weaver10} founded two important results for BAT-detected AGNs: 1) there exists a very tight relationship between [OIV]25.89\micron~and another important IR emission line: [NeV]14\micron, and 2) the [NeV]14\micron~is formed by photoionization from the AGN, since very high-energy photons are required in order to form the line (its ionization potential, IP, is 97.2~eV; see also \citealt{pereira10,davies17}). Based on these two points they showed and concluded that the emission of [OIV]25.89\micron~ (whose IP is almost half of that of [NeV]) in BAT sources also responds to photoionization by the AGN.

So, according the AGN unification model \citep{urry95}, [OIII]$\lambda5007$\AA~and [OIV]25.89\micron~emission-lines originate in the photoionized gas that forms the NLR. The same powerful radiation field provides the ionizing photons required to form both [OIII]$\lambda5007$\AA~and [OIV]25.89\micron. While the optical [OIII]$\lambda5007$\AA~ emission line is {\it the} widely accepted proxy of the AGN primary continuum \citep[e.g.][]{bassani99,schmitt03,heckman05,lamassa10,zhang17}, in the case of the IR [OIV]25.89\micron~ emission line there has been a long standing debate trying to clarify if both its existence and intensity are related to the AGN or to the intense starburst continuum \citep[e.g.][]{dale06,hao09}. However, it is now accepted that, albeit collisional ionization may play an important role in some cases, the AGN is the main responsible for the excitation of such a line, specially in the case of the bright AGN constituting the bulk of the extragalactic population in the BAT catalogue \citep{lamassa10,weaver10}. In this sense, we highlight that 24 out of 35 objects where OVII~(f) or OVIII~Ly$\alpha$ were confidently measured, are BAT sources (69\% of the sub-sample). The [OIII] data were collected from \citet{schmitt03,heckman05,gu06,panessa06,bentz09,schles09,greene10}, while [OIV] data were collected from \citet{schweitzer06,deo07,diamond09,pereira10,sales10,tommasin10,weaver10,dasyra11,wu11}.

%...... MIR-12micron
Another well-known proxy of the AGN is the MIR 12\micron~continuum emission; its strong relationship with the hard X-ray 2--10~keV band \citep{gandhi09} makes it another indicator of the AGN power. According to the unified scenarios for AGNs, large amounts of gas and dust surround the central region. Therefore, their temperatures raise up by absorption of high energy photons (UV, X-rays) of the primary continuum. The warm dust is mixed within the circumnuclear ionized gas and thus is mainly affected by the nuclear ionizing radiation and also by resonant-scattered Ly$\alpha$ photons created (and trapped) inside the nebula \citep{telesco88,mouri92}. Once the dust reaches the equilibrium temperature it re-radiates in the MIR, reaching the maximum emission among 7--26\micron~\citep{pier92,ramos09,asmus14}. MIR 12\micron~ continuum radiation is emitted by warm gas (200--600~K) after it absorbs the primary continuum. Lower-temperature ($\sim$10--100~K) gas also emits in the Far Infrared ($\lambda\gtrsim$40\micron; FIR) band, but the main contributors to this band are star-forming regions, as it will be discussed later. The 12\micron~ continuum has been pointed out as good tracer of warm-hot dust in the AGN scenario \citep{horst06,gurkan15}. 

One of the most relevant advantages of using the MIR is the high angular resolution with which data are obtained. Nowadays, ground-based 8m-class telescopes allow us to reach arcsec and/or diffraction-limited resolution \citep{krabbe01}. On-orbit telescopes, as {\it ISO} and {\it Spitzer}, have been great instruments mainly due to their high sensitivity but they do not reach sub-arcsec resolution \citep{asmus14}. \citet{horst09} showed that 40\% of Spitzer's data are contaminated by extended emission in comparison with VLT/Visir's data. So, having sub-arcsec resolution is particularly important in the case of AGN in the local Universe because it is crucial to isolate the nuclear emission (dust mixed within circumnuclear ionized gas) as much as possible. Taking this into account, we are using the most recent collection of 12\micron~data \citep{asmus14}, obtained with state-of-the-art ground-based IR instruments: Visir from VLT, Comics from Subaru and Michelle and T-ReCS (yet retired instruments) from Gemini.

%%%%%%%%%%%%%%%%%%%%%%%%%%%%%%%%%%%%%%%%%%%%%%%%%%%%%%%%%%%%%%%%%%%%%%%%%%%%%%%%%%%%%%%%%%%%%%%
\begin{table}\label{instruments}
\caption{List of telescopes and instrumets used in this {\it paper}.}\label{instruments}
\begin{tabular}{|l|c|}
\hline
Telescope & Instrument \\
\hline
 HST & WFPC2  \\
     & WF/PC1 \\
     & FOC    \\

La silla (1.52m) & B\&C \\
La silla/MPI (2.2m) & B\&C \\
La silla (3.6m)  & EFOSC \\

Palomar/Hale(5m) & double spec.\\

Lick/Shane (3m) & Kast dual spec. \\

Las Campanas/du Pont (2.5m) & B\&C \\
Las Campanas/APO (3.5m)     & DIS \\

Subaru & COMICS \\

Gemini & Michelle  \\
       & T-ReCS    \\
VLT    & VISIR     \\
Spitzer & IRS \\
IRAS &  \\
Swift & BAT  \\
XMM-Newton & EPIC  \\
           & RGS   \\
\hline
\end{tabular}
\end{table}
%%%%%%%%%%%%%%%%%%%%%%%%%%%%%%%%%%%%%%%%%%%%%%%%%%%%%%%%%%%%%%%%%%%%%%%%%%%%%%%%%%%%%%%%%%%%%%%%%%%%%%%%%

The FIR continuum emission, on the other hand, is produced by the very cold gas surrounding star-forming regions.The FIR 60\micron~and 100\micron~data (IRASC 1988, The Point Source Catalog, Version 2.0, \citealt{sanders89,moshir90,sanders03,surace04,lisenfeld07,serje09}) will be used as proxies of the SB activity \citep{rodriguez86,rodriguez87,mouri92,strickland04,hatzi10}.

Table~\ref{instruments} shows the facilities (telescopes and instruments) that were employed to obtain the data we are using along this {\it paper}.

%%%%%%%%%%%%%%%%%%%%%%%%%%%%%%%%%%%%%%%%%%%%%%%%%%%%%%%%%%%%%%%%%%%%%%%%%%%%%%%%%%%
%
% 			RESULTS 
%
%%%%%%%%%%%%%%%%%%%%%%%%%%%%%%%%%%%%%%%%%%%%%%%%%%%%%%%%%%%%%%%%%%%%%%%%%%%%%%%%%%%

\subsection{An IR diagnostic diagram for obscured AGN}\label{DD}

We are interested in using [OIV] as a proxy of the AGN, but we can be sure about that property only for 69\% of our sample, which constitutes the sub-sample of BAT sources within CHRESOS. This section is devoted to make a last test over [OIV]25.89\micron~in order to show that this line {\it is} a proxy of the AGN for the whole CHRESOS's sources, regardeless their BAT or non-BAT condition.

\begin{figure*}
\centering
\includegraphics[angle=270,width=0.85\textwidth]{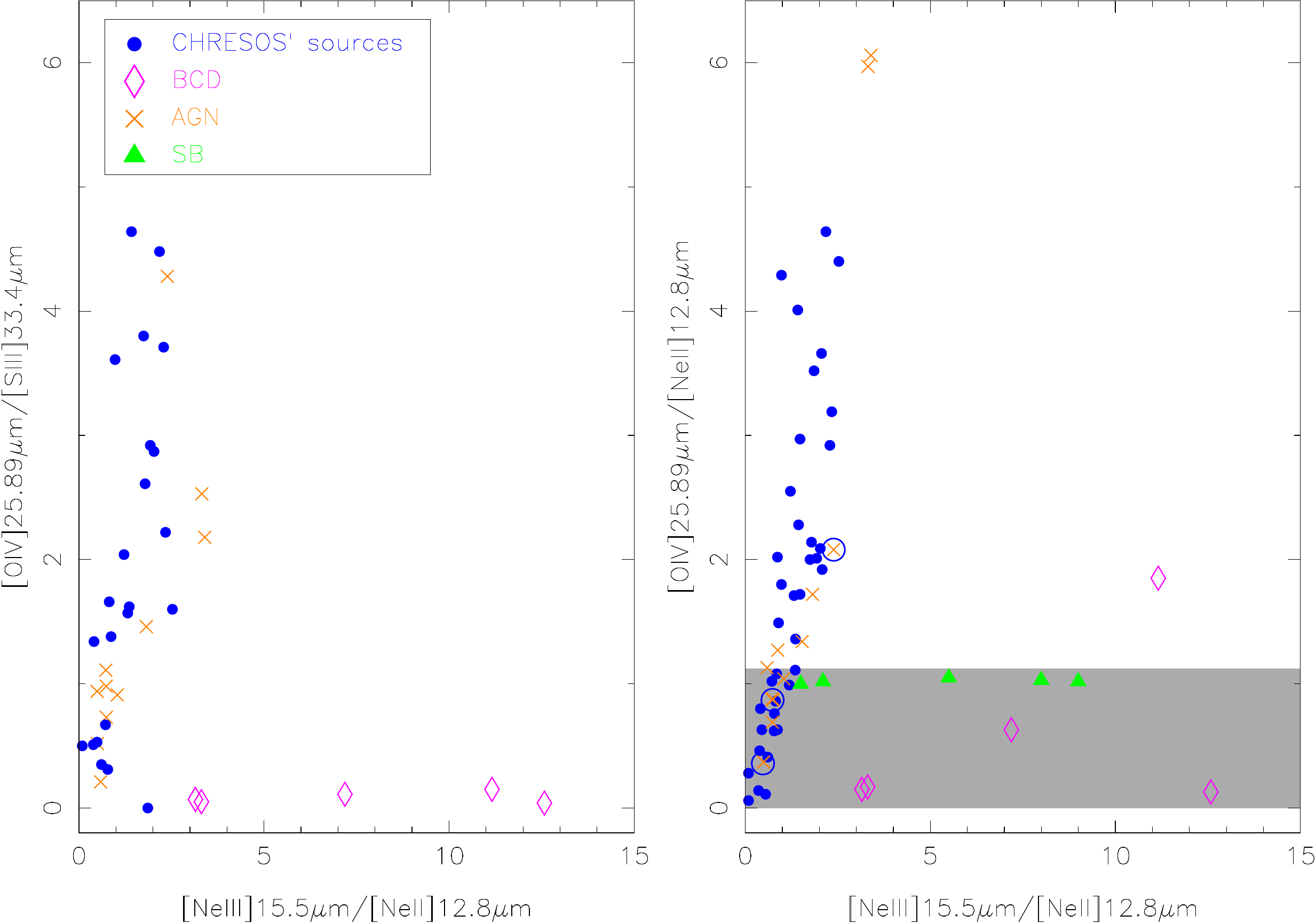}
\caption{IR diagnostic diagrams aiming at separating AGNs from Blue Compact Dwarf galaxies \citep{hao09}. The AGN-branch is the almost vertical one in the two diagrams. Most CHRESOS sources are distributed along the AGN branch. The sources located within the grey rectangle show an important contribution from star forming regions.}\label{F:IR-BPT}
\end{figure*}

Figure~\ref{F:IR-BPT} shows IR-BPT diagnostic diagrams, which were developed by \citet{hao09} with the aim of distinguishing the [OIV]25.89\micron~emission from AGN and the [OIV]25.89\micron~emission from Blue Compact Dwarf (BCD) galaxies. The diagnostic power of these diagrams rely on the high-energy photons (54.94~eV) required to form the [OIV]25.89\micron: the radiation field of any Main-Sequence-central-star(s) HII region is not strong enough to account for the [OIV] observations. Instead, the harder ionizing field of Wolf Rayet (WR) stars is able to account for [OIV] (and other high excitation/ionization) line emission, and this is what makes BCD galaxies so relevant: they are extreme SB, suspected to contain a significant number of WR stars \citep{lutz98,hao09,weaver10,burtscher15}. The diagrams of Fig.~\ref{F:IR-BPT} are formed with MIR emission-line ratios: [OIV]25.89\micron/[SIII]33.4\micron~vs. [NeIII]15.5\micron/[NeII]12.8\micron~(left-hand diagram) and [OIV]25.89\micron/[NeII]12.8\micron~vs. [NeIII]15.5\micron/[NeII]12.8\micron~(right-hand diagram). \citet{hao09} have demonstrated that, in effect, BCDs and AGNs tend to be located in different zones because of [OIV]25.89\micron~being powered by different mechanisms. We have plotted:\\

\begin{itemize}

\item CHRESOS' sources as (blue) filled circles (the IR data of CHRESOS's sources were obtanied from \citealt{weaver10,gallimore10});
\item AGNs from \citet{sales10} as (orange) crosses; they are used as a control sample. The encircled sources also belong to CHRESOS;
\item SB galaxies from \citet{lutz96,lutz98} as (green) filled triangles; in addition, SB from \citet{genzel98} are not shown individually but globally as the grey rectangle (shaded area);
\item BCDs as examples of ``extreme starburst" \citep{hao09}; they are the (pink) empty diamonds. We assess their location within the diagrams in comparison to AGNs and ``normal'' SB.

\end{itemize}

The diagrams actually separate AGNs from BCD galaxies in two branches (being the AGN branch the almost vertical one), but both of these branches also contain {\it normal} SB galaxies. They are equivalent to the optical BPT diagnostic diagram ([OIII]$\lambda5007$/H$\beta$ vs. [NII]$\lambda6583$/H$\alpha$; \citealt{baldwin81,veilleux87}) in the sense that there exists a correspondence among the ``pure AGN'' zone, intermediate zone, and SB-dominated BCD zone both in the IR and in the optical diagram according to \citet{K01,K03}, and \citet{hao09}. The sources for which [OIV]25.89\micron~/[SIII]33.4\micron~$\sim1-10$ (left-hand diagram) are clearly dominated by the AGN emission, and no SB is located in that zone (the ``pure AGNs''); however, the two branches seem to have a common origin \citep[vertix; see the fig.~1 of][]{hao09}, and many SB with and without WR contribution are located in that zone (each group tending to follow their respective branch). The distributions of sources in the two diagrams of Fig.~\ref{F:IR-BPT} are very similar \citep{hao09}. As there are more [NeII]12.8\micron~data than [SIII]33.4\micron~data in the literature for both CHRESOS' sample and comparison samples, that equivalence is particulary useful in our case.

We observe the very same behaviour both in CHRESOS' sources and in the control sample of \citet{sales10}: all these sources are located along the (vertical) AGN branch. As expected, a significant fraction of our sample (37\%) is in the parameter space that is shared between AGN and SB. Nonetheless, we emphasize that none of our sources show any kind of signature of WR stars accounting for the [OIV] line-emission. The result of the current analysis is twofold. First, it reasserts the AGN nature of CHRESOS's sources, which was expected since the hard X-ray luminosity for most of the sample is just too high ($L_{2-10~keV}\gtrsim 10^{42}$~erg~s$^{-1}$) not to harbor an AGN. And, on the other hand, it also confirms that [OIV]25.89\micron~is a {\it bona fide} indicator of the AGN power in CHRESOS' sources, independently of their BAT/non-BAT condition.

\section{Model predictions.}
\label{sect:predictions}

Ionizing power is very different in AGN and in SB. However, both the AGN and the SB scenarios are capable of producing soft X-ray line emission. Two distinct processes are involved in each case: photoionization (PIE) and collisional ionization/excitation (CIE). In this Section, we show PIE and CIE predictions onto OVIII~Ly$\alpha$/OVII~(f) line-ratio in order to compare them with our own observations.

%PIE

\begin{figure}
\centering
\includegraphics[angle=0,width=0.45\textwidth]{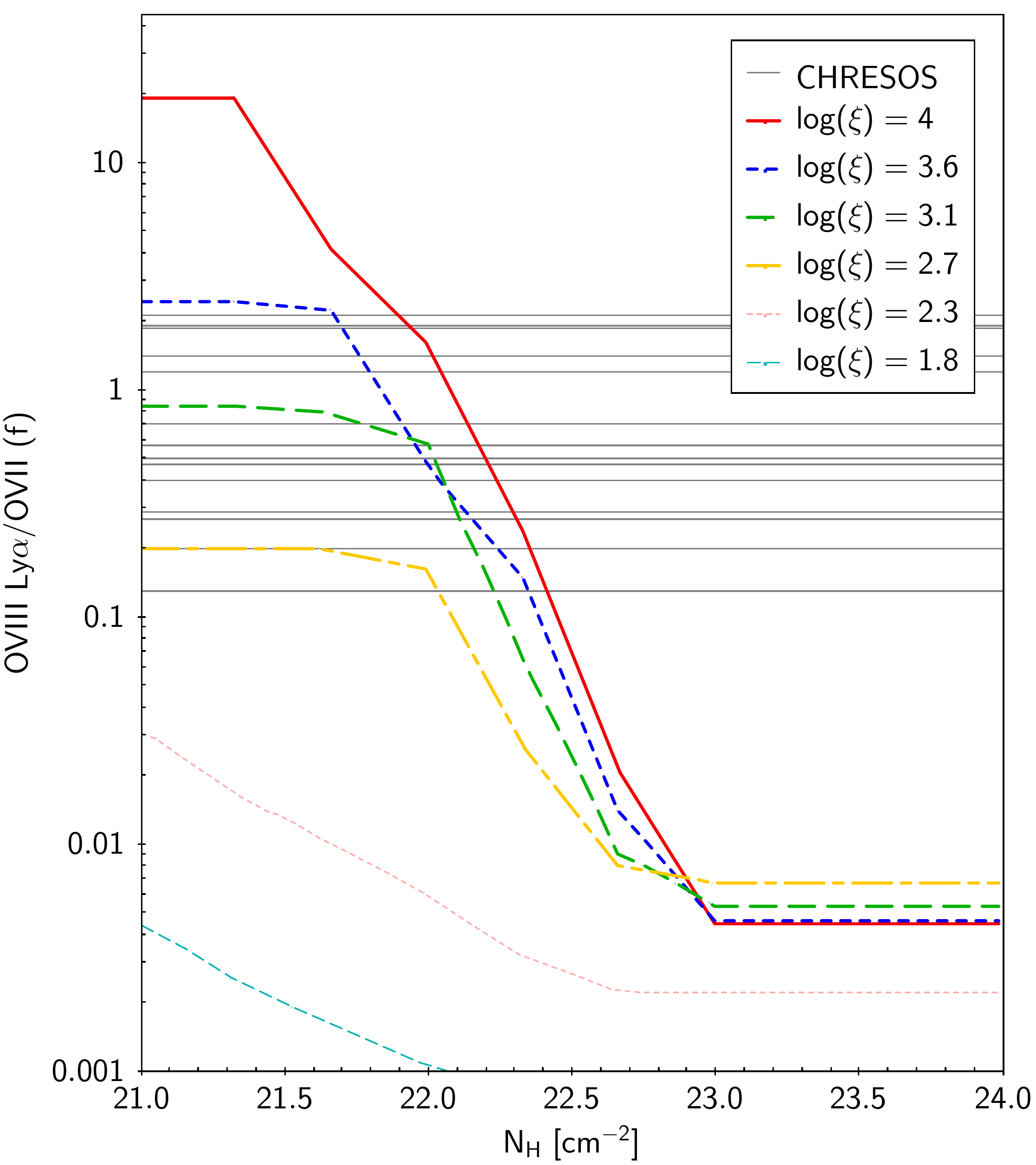}
\caption{XSTAR model prediction on the OVIII~Ly$\alpha$/OVII~(f) flux ratio as a function of column density (N$_{\rm H}$) for different values of the ionization parameters. The thin (grey) horizontal lines represent our sources (see also Fig.~\ref{F:CIE}).}\label{F:PIE}
\end{figure}

PIE gathers two important processes: radiative decay after continuum photoexcitation and recombination plus radiative cascades after photoionization. Their relative weigth is governed by the column density ($N_{\rm H}$). The former dominates at low $N_{\rm H}$, the latter dominates at high $N_{\rm H}$ \citep{kink02}. \citet{kallman14} performed a very-accurate spectral fitting analysis to the soft X-ray ACIS/{\it Chandra} spectra of NGC~1068 by using a version of the XSTAR code, which incorporates the continuum excitation process. We have used this same separate version of XSTAR \citep{kallman01} to compare the observed OVIII~Ly$\alpha$/OVII~(f) line ratios in CHRESOS with its predictions. To do so, we have modelled a set of constant-density ($n_e=10^4$~cm$^{-3}$) nebulae irradiated by an AGN-like radiation field. We have adopted the default solar abundances in XSTAR. The shape of the incident continuum between 0.1~eV -- 200~keV was built according to the AGN Spectral Energy Distribution (SED) after \citet{korista97-apjs}, taking into account both the high-energy power-law and the UV Big-Blue Bump as follows:

$f_{\nu}=\nu^{\alpha(UV)}exp(-h\nu/kT_{BB})exp(-kT_{IR}/h\nu)+a.\nu^{\alpha_X}$

with $\alpha(UV)=-0.5$; $kT_{BB}=30$~eV; $kT_{IR}=0.1$~eV and $\alpha_X=-0.7$. A grid of models was obtained by varying the column density in the range $10^{21}$~cm$^{-2}<N_{\rm H}<10^{24}$~cm$^{-2}$ and varying the ionization parameter $\xi$ in the range $0<log(\xi)<4$ (the ionization parameter is defined as $\xi=L_{\rm ion}/n_Hr^2$, where $L_{\rm ion}$ is the luminosity of the ionizing source, integrated from 13.6~eV, the H-ionizing energy; $n_H$ is the hydrogen number density, and $r$ is the distance from the gas cloud to the ionizing source). Changes in the ambient density (in the range $n_e \lesssim 10^{10}cm^{-3}$, as stated by \citealt{kink02}) do not change substantially the model predictions. Our model predictions onto the OVIII~Ly$\alpha$/OVII~(f) ratio are shown in Figure~\ref{F:PIE}. The observed line ratios in CHRESOS are shown as grey, thin continuus lines.

PIE models fully reproduce the observed OVIII~Ly$\alpha$/OVII~(f) ratio as long as the ionization parameter is in the range $2.7<log(\xi)<4$ and the column density remains lower than $10^{22}$~cm$^{-2}$, in agreement with previous results of \citet{kink02} and \citet{kallman14}. 

%CIE

The nuclear SB, when present, can become important sources of X-rays too; they contribute to the X-ray spectra through different processes. The interstellar medium (ISM) can be heated up to very high temperatures ($T\sim10^6$~K) due to winds of massive stars and supernovae (SN) explotions. These processes account for thermal-plasma emission, which mainly emerges in the soft X-ray band \citep{persic02,levenson05,masHesse08}. But the bursts of star formation also give raise to accreting binary systems which are also sources of hard X-rays. While the latter processes are responsible for a power-law-type harder spectrum, the former produce a line-emission-dominated spectrum similar to (but still distinguishable in many aspects from) that of PIE \citep{kink02}.

We utilized XSPEC v12.9.1p and {\sc apec} v3.0.7 \citep{arnaud96} in order to obtain the line ratios. We used XSPEC {\sc flux} command to calculate the fluxes of each line with respect to the underlying continuum model. For OVIII~Ly$\alpha$ we considered the 0.647--0.661 band, and for OVII~(f) the 0.556--0.566~keV band. We constructed a grid of {\sc apec} models considering 30 logarithmic steps in the plasma temperature: $0.04<kT<1.2$~keV. To obtain the underlying continuum level, we calculated the flux of the continuum model by setting the abundance parameter to zero. Then, we fixed the abundance to 30 logarithmic steps in the 0.01--4 range and we calculated the ratios by subtracting the flux obtained with the abundance fixed to zero as follows: $[F_2(kT,Z)-F_2(kT,Z=0)]/[F_1(kT,Z)-F_1(kT,Z=0)]$. By doing this, we observed that the line ratios were always independent of the abundance, as expected.

\begin{figure}
\centering
\includegraphics[angle=0,width=0.45\textwidth]{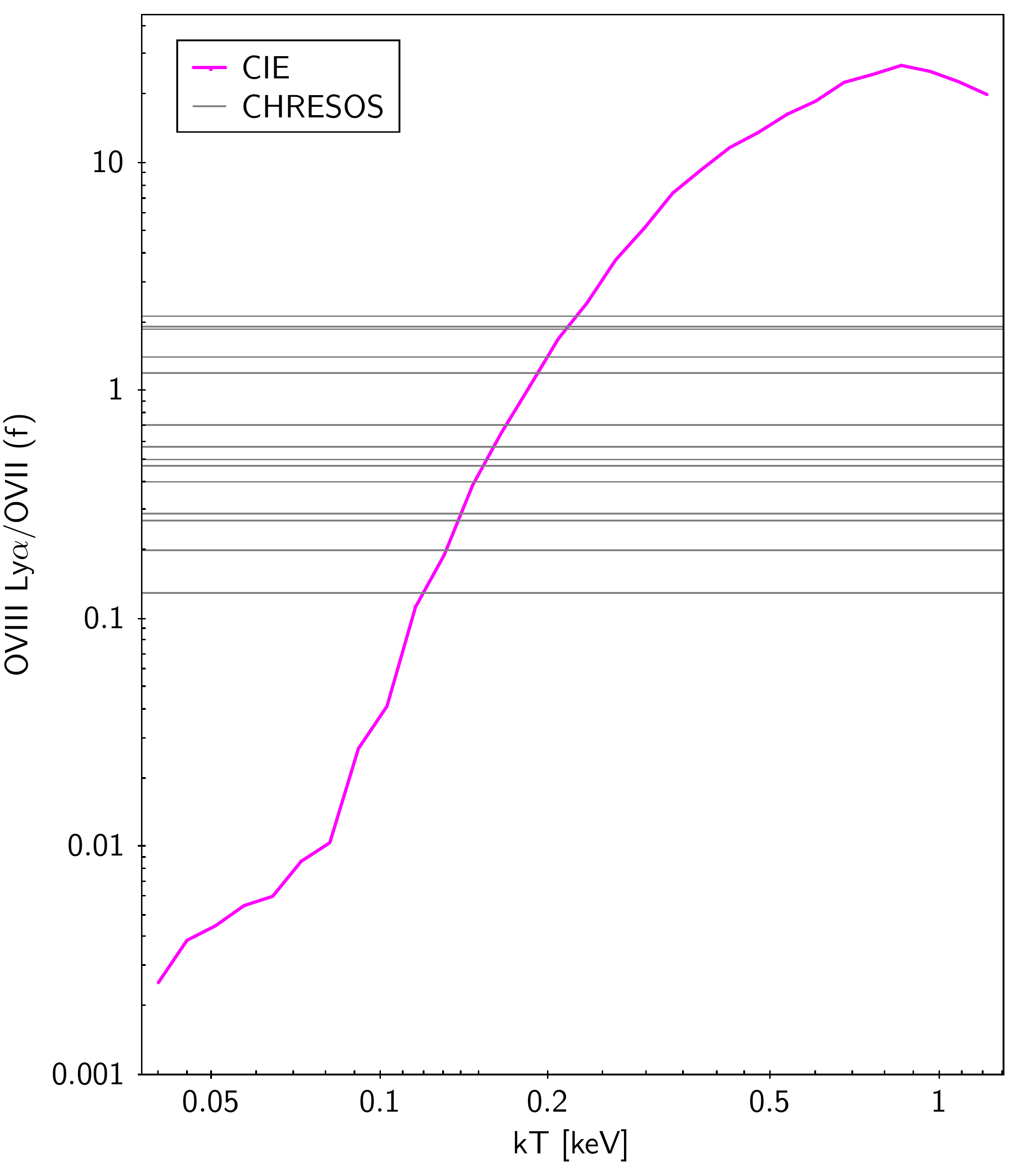}
\includegraphics[angle=270,width=0.4\textwidth]{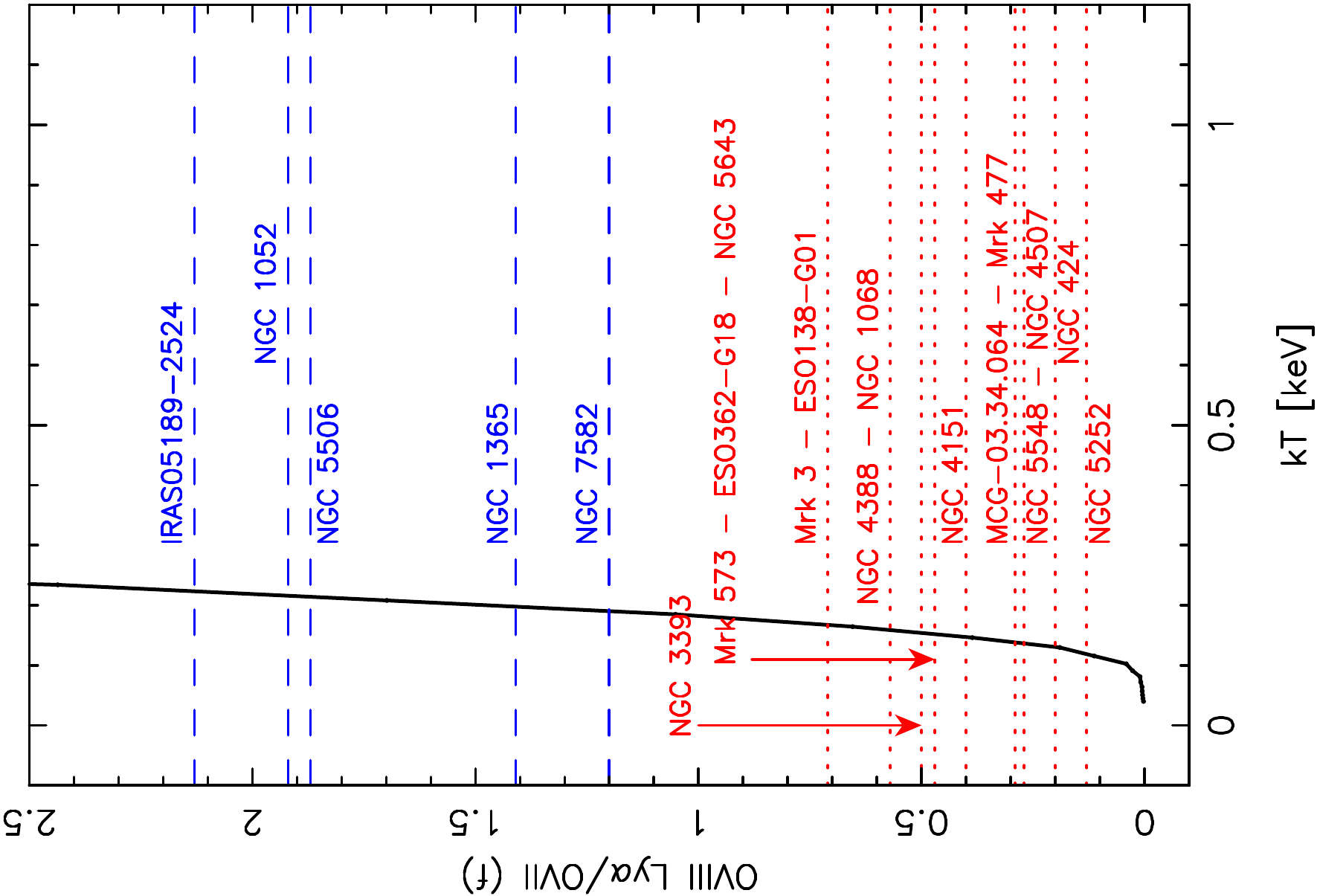}
\caption{{\it Top:} CIE ({\it apec}) model prediction on the OVIII~Ly$\alpha$/OVII~(f) flux-ratio. {\it Bottom:} CHRESOS's sources shown individually according to their observed line-ratios. The highest ratios (OVIII~Ly$\alpha$/OVII~(f) $>1$) are shown in (blue) dashed lines, while the lowest ratios are shown in (red) dotted lines. The continuus line represents the CIE model of the top panel.}\label{F:CIE}
\end{figure}

The results of CIE models are shown in the top panel of Figure~\ref{F:CIE}. The OVIII~Ly$\alpha$/OVII~(f) flux ratio grows up while the gas temperature ($kT$) increases from $0.1$~keV to $\sim0.8$~keV. Most of the sources present OVIII~Ly$\alpha$/OVII~(f)$<1$, which means that collisional gas should have temperatures lower than $0.15$~keV. Other few sources (5) show OVIII~Ly$\alpha$/OVII~(f) up to 2.12, requiring a fitting-temperature raising up to $kT\lesssim 0.25$~keV. This is illustrated in the bottom panel of Fig.~\ref{F:CIE}, where we have shrinked the parameter space to the observed range of OVIII~Ly$\alpha$/OVII~(f). Sources are shown individually over its own line ratio; the lowest (highest) ratios are shown in dotted (dashed) lines. According to the empirical criterion of \citet{guainazzi07}, SB-contribution could be relevant for sources with OVIII~Ly$\alpha$/OVII~(f)$>1$, since this value represents (with an additional constraint on the total oxygen luminosity) the level at which SB could dominate the soft X-ray spectra. Hence, CIE models are able to reproduce all of our observations; there exists a set of parameters that can explain the observed data.

Our simple modeling of the PIE and CIE processes can predict the observed OVIII~Ly$\alpha$/OVII~(f) line ratios. These results by themselves, however, are not enough to disentangle which is the main mechanism that powers the X-ray oxygen line-emission. 

%%%%%%%%%%%%%%%%%%%%%%%%%%%%%%%%%%%%%%%%%%%%%%%%%%%%%%%%%%%%%%%%%%%%%%%%%%%%%%%%%%%

%%%%%%%%%%%%%%%%%%%%%%%%%%%%%%%%%%%%%%%%%%%%%%%%%%%%%%%%%%%%%%%%%%%%%%%%%%%%%%%%%%%
%
% 			DISCUSSION 
%
%%%%%%%%%%%%%%%%%%%%%%%%%%%%%%%%%%%%%%%%%%%%%%%%%%%%%%%%%%%%%%%%%%%%%%%%%%%%%%%%%%%

\section{Multi-wavelength correlations}\label{AGN}

In the following, we are going to test OVII~(f) and OVIII~Ly$\alpha$ against proxies of the AGN and the SB ionizing power throughout correlations between luminosities. We have studied both flux- and luminosity-diagrams, and have discarded flux-correlation analysis because, in our very limited redshift range, all the corresponding relationships are heavily influenced by two sources that are significantly brighter than the others. They are the Seyfert~2 NGC~1068 and the Seyfert~1.5 NGC~4151. Their observed fluxes (both in OVII~(f) and OVIII~Ly$\alpha$) are almost two orders of magnitude higher than those for the remaining sample. In Appendix~\ref{test} we will show that the luminosity correlations are not affected by the distance.

\subsection{X-ray continuum photometry}

%...... X-Ray continua

\begin{figure*}
\centering
\includegraphics[angle=270,width=0.8\textwidth]{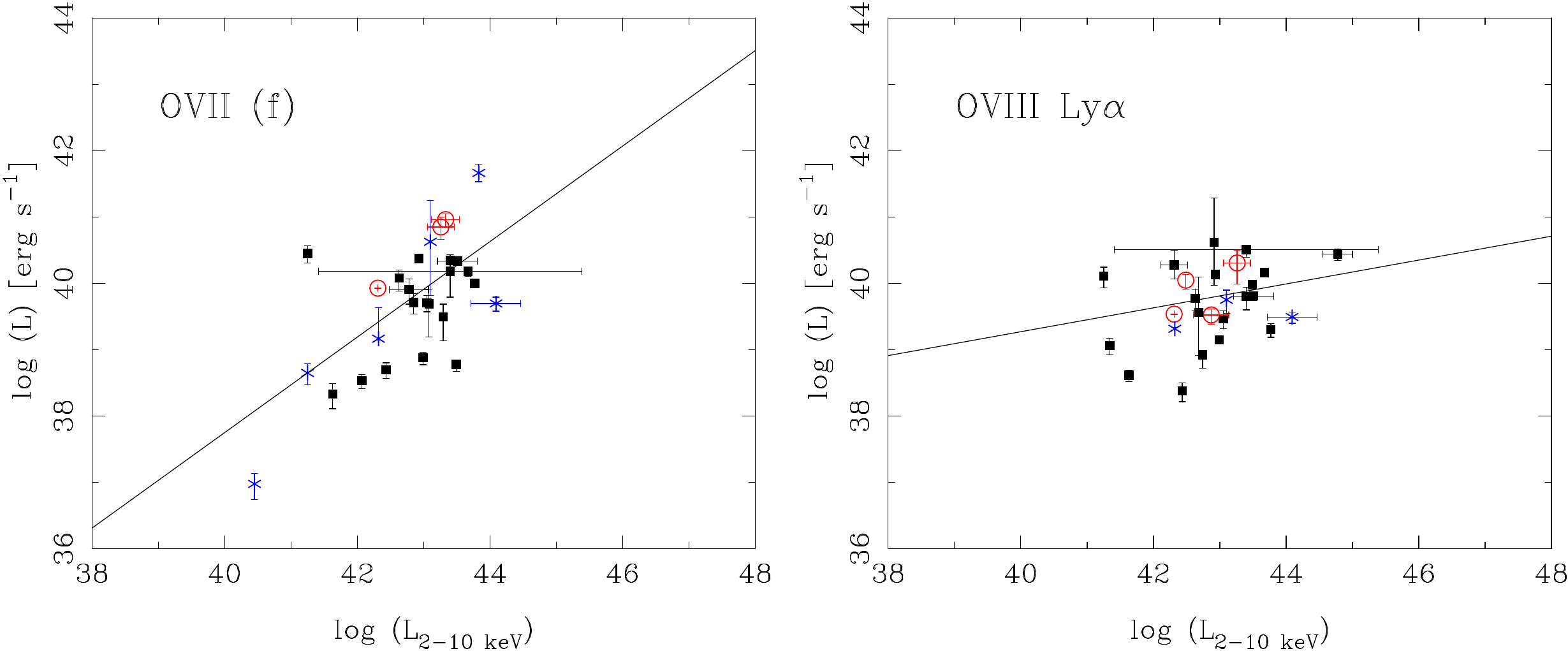}
\includegraphics[angle=270,width=0.8\textwidth]{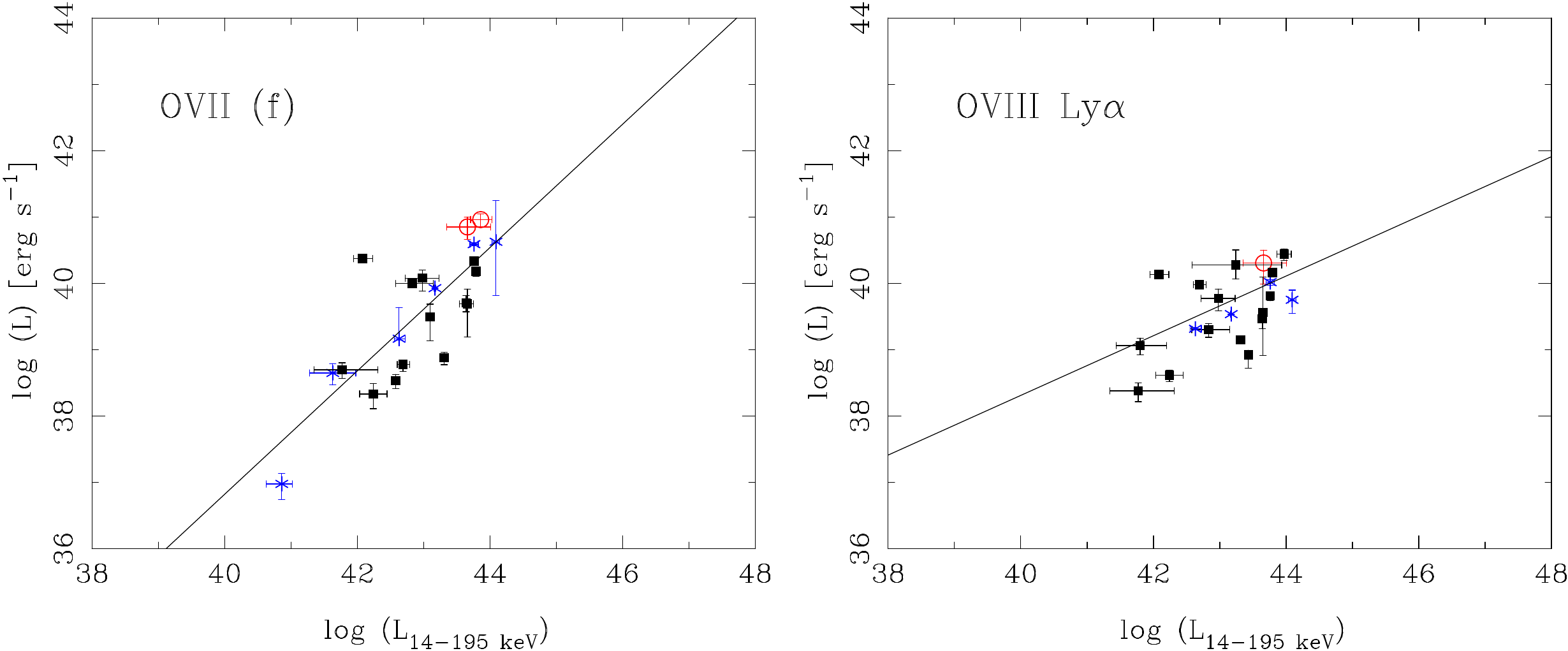}
\caption{X-ray oxygen emission lines OVII~(f) and OVIII~Ly$\alpha$ vs. the two primary, hard X-ray continuum bands: 2--10~keV (upper diagrams), 14--195~keV (lower diagrams). Sources are described as follows: Seyfert~1-1.2 sources are plotted as empty (red) circles; Seyfert~1.5-1.9 as (blue) asterisks, and Seyfert~2 as filled (black) squares; unclassified sources were drawn as empty (orange) diamonds.. The continuous line represents the linear fitting.}\label{F:LvsC_HX}
\end{figure*}

In Figure~\ref{F:LvsC_HX} we compare the luminosity of OVII~(f) and OVIII~Ly$\alpha$ with that of the primary continuum bands: 2--10~keV and the BAT band (14--195~keV). Hereafter, the sources are ordered in groups according to their Seyfert sub-type: Seyfert~1-1.2 sources are plotted as empty (red) circles; Seyfert~1.5-1.9 are drawn as (blue) asterisks, and Seyfert~2 as filled (black) squares; not-yet-classified sources (as stated by NED) were drawn as empty (orange) diamonds. 

We have chosen to work with two hard X-ray continuum bands with the aim of having direct, different, and complementary measurements of the primary continuum. Below 10~keV the central X-ray source in Seyfert~2 objects can be severely attenuated if the intervening column density is $N_{\rm H}>10^{22.5}$~cm$^{-2}$. Intermediate gas density ($N_{\rm H}<1.5\times10^{24}$~cm$^{-2}$) let the radiation above $\sim10$~keV penetrate the torus and reach the observer \citep{turner97}, but if the density is $N_{\rm H}\gtrsim 3\times10^{24}$~cm$^{-2}$ (the limit sets the Compton-thick, CT, regime) the energy above 10~keV, {\it i.e.} the emission from the BAT band, will also be blocked \citep{melendez08,weaver10}. We have a few CT sources in our sample, according to the last results of \citet{baumg13,ricci15} and \citet{oh18}. In order to be conservative (2--10~keV luminosities from the literature are indeed absorption-corrected), we have used the two bands. So even if the 2--10~keV luminosities were attenuated, the 14--195~keV luminosities would remain unaffected for most of the sample unless the column density is $\ge$10$^{25}$~cm$^{-2}$. As long as our analysis is concerned, the CT sources do not occupy a specific locus in the 14--195~keV luminosity parameter space.

\begin{table*} 
\centering
\caption{Linear fitting parameters, normalized to $L=10^{40}$~erg$\cdot$ s$^{-1}$.}\label{T:correlation}
\label{T:regresionsLC}
\begin{tabular}{lcccccc}
\hline\hline
Continuum band/Proxy & Emission Line & slope & intercept & $\rho_S$ & {\it p-value} & Number of data points \\
\hline
2--10~keV  & OVII (f) & $0.72\pm0.17$ & $37.75\pm 0.49$ & 0.45 & $1.2\times10^{-2}$ & 31 \\ 

	      & OVIII Ly$_\alpha$ & $0.31\pm 0.13$ & $39.22\pm 0.36$ & 0.22 & $2.4\times10^{-1}$ & 29\\ 

\hline
14--195~keV & OVII (f) & $0.93\pm 0.16$ & $36.81\pm 0.49$ & 0.77 & {\boldmath $5\times10^{-5}$} & 21\\ 

           & OVIII Ly$_\alpha$ & $0.45\pm 0.16$ & $38.2\pm 0.49$ & 0.55 & $1.18\times10^{-2}$ & 20\\ 
\hline\hline
[OIII]$\lambda5007$ & OVII (f) & $0.61\pm 0.11$ & $39.28\pm 0.14$ & 0.54 & {\boldmath $3.29\times10^{-3}$} & 28\\ 

		            & OVIII Ly$\alpha$ & $0.3\pm 0.11$ & $39.49\pm 0.13$ & 0.48 & $1.46\times10^{-2}$ & 25\\
\hline
[OIV]25.89\micron & OVII (f) & $ 0.75\pm 0.15$ & $38.94\pm 0.21$ & 0.45 & $2.29\times10^{-2}$ & 25\\

		          & OVIII Ly$\alpha$ & $0.53\pm 0.12$ & $39.1\pm 0.16$ & 0.68 & {\boldmath $2.7\times10^{-4}$} & 24\\ 
\hline
MIR-12\micron & OVII (f) & $0.68\pm 0.12$ & $37.59\pm 0.37$ & 0.62 & {\boldmath $1.26\times10^{-3}$} & 24\\ 

	          & OVIII Ly$\alpha$ & $0.57\pm 0.14$ & $37.83\pm 0.45$ & 0.61 & $2.14\times10^{-2}$ & 23 \\ 
\hline\hline
60\micron & OVII (f) & $0.74\pm 0.19$ & $37.04\pm 0.68$ & 0.44 & $2.07\times10^{-2}$ & 28\\ 

	      & OVIII Ly$\alpha$ & $0.42\pm 0.13$ & $38.06\pm 0.49$ & 0.51 & {\boldmath $9.45\times10^{-3}$} & 25\\ 
\hline
100\micron & OVII (f) & $0.75\pm 0.22$ & $36.87\pm 0.79$ & 0.58 & $1.18\times10^{-2}$ & 18\\ 

	       & OVIII Ly$\alpha$ & $0.46\pm 0.15$ & $37.92\pm 0.57$ & 0.55 & $1.75\times10^{-2}$ & 18\\
\hline
\end{tabular}
\end{table*}

Linear fittings were performed in order to quantitavely analyze putative relationships among the luminosities. These and all other regression and correlation parameters are listed in Table~\ref{T:correlation}: slopes and intercepts of every regression line; Spearman correlation coefficient, $\rho_S$; {\it p-value}, as a measurement of the statistical significance; and the sample size in each diagram. Data distribution and Spearman correlation coefficients suggest that the X-ray oxygen emission lines are more strongly correlated with the BAT band than with the 2--10~keV band. Note that the fitting slopes with the 2--10~keV and 14--195~keV bands are very different too, meaning that no single relationship applies to both of them. 

According to the {\it p-value}, the relationship between OVII~(f) and the hardest X-ray band (14--195~keV) is statistically significant (99\% confidence level). In fact, this is the strongest relationship we have found, out of 5 statistically significant relationships (boldfaced in Table~\ref{T:correlation}). We have also ruled out that the correlation is governed by the distance by performing the so-called Scrambling Test (Appendix~\ref{test}). We obtain a probability lower than 1\% that the observed trend is driven by the distance.

%%%%%%%%%%%%%%%%%%%%%%%%%%%%%%%%%%%%%%%%%%%%%%%%%%%%%%%%%%%%%%%%%%%%%%%%%%%%%%%%%%%%%%%%%%%%%%%%%%%%%%%%%%%%%%%%%%%%%%%%%%%%%%%%%%%%%%%%%%%%%%%%%%%%%%%%%%%%%%%%%%%%%%%
\subsection{IR/optical AGN spectroscopic indicators}
\label{NLR}

\begin{figure*}
\centering
\includegraphics[angle=270,width=0.8\textwidth]{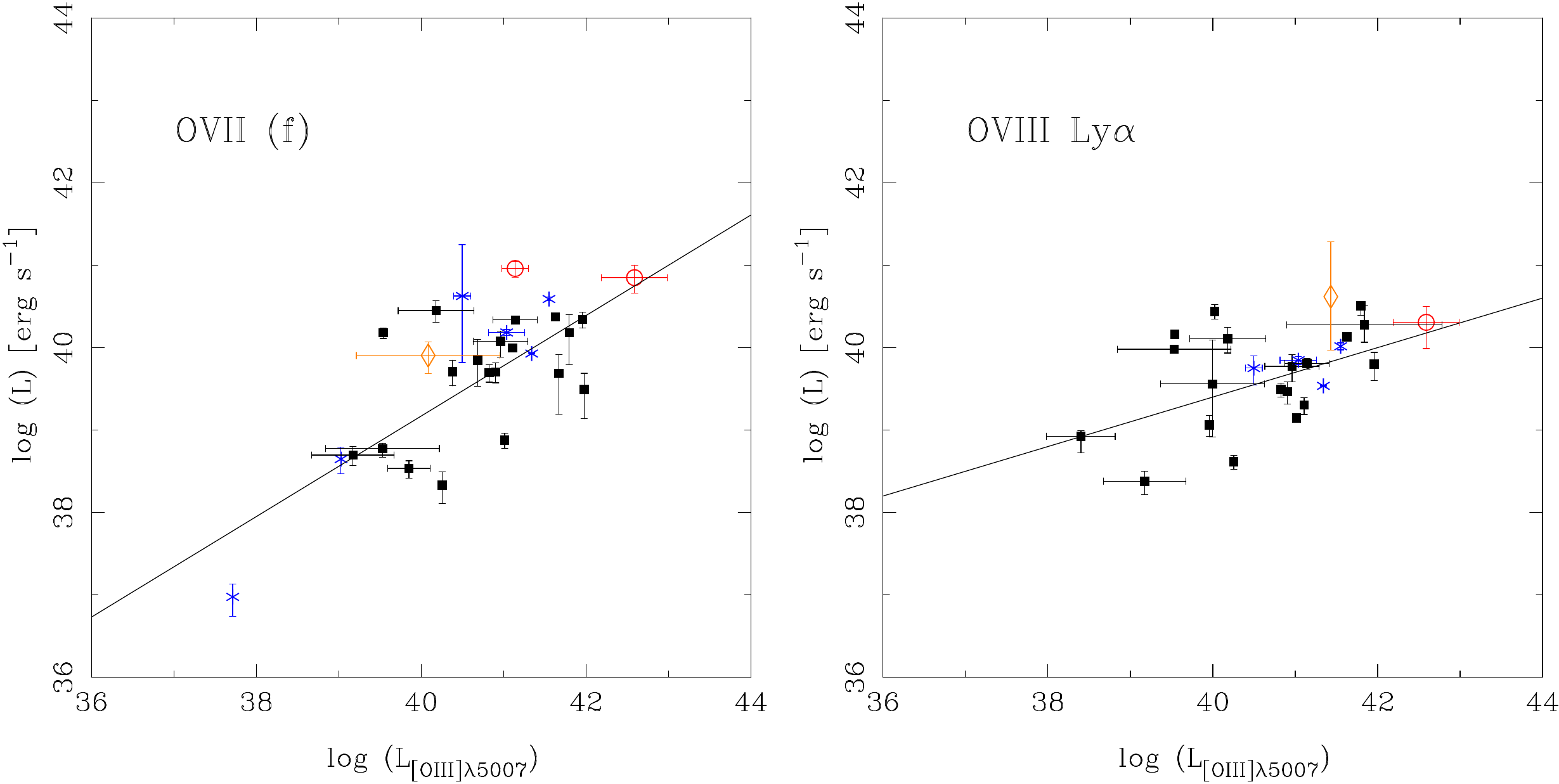}
\includegraphics[angle=270,width=0.8\textwidth]{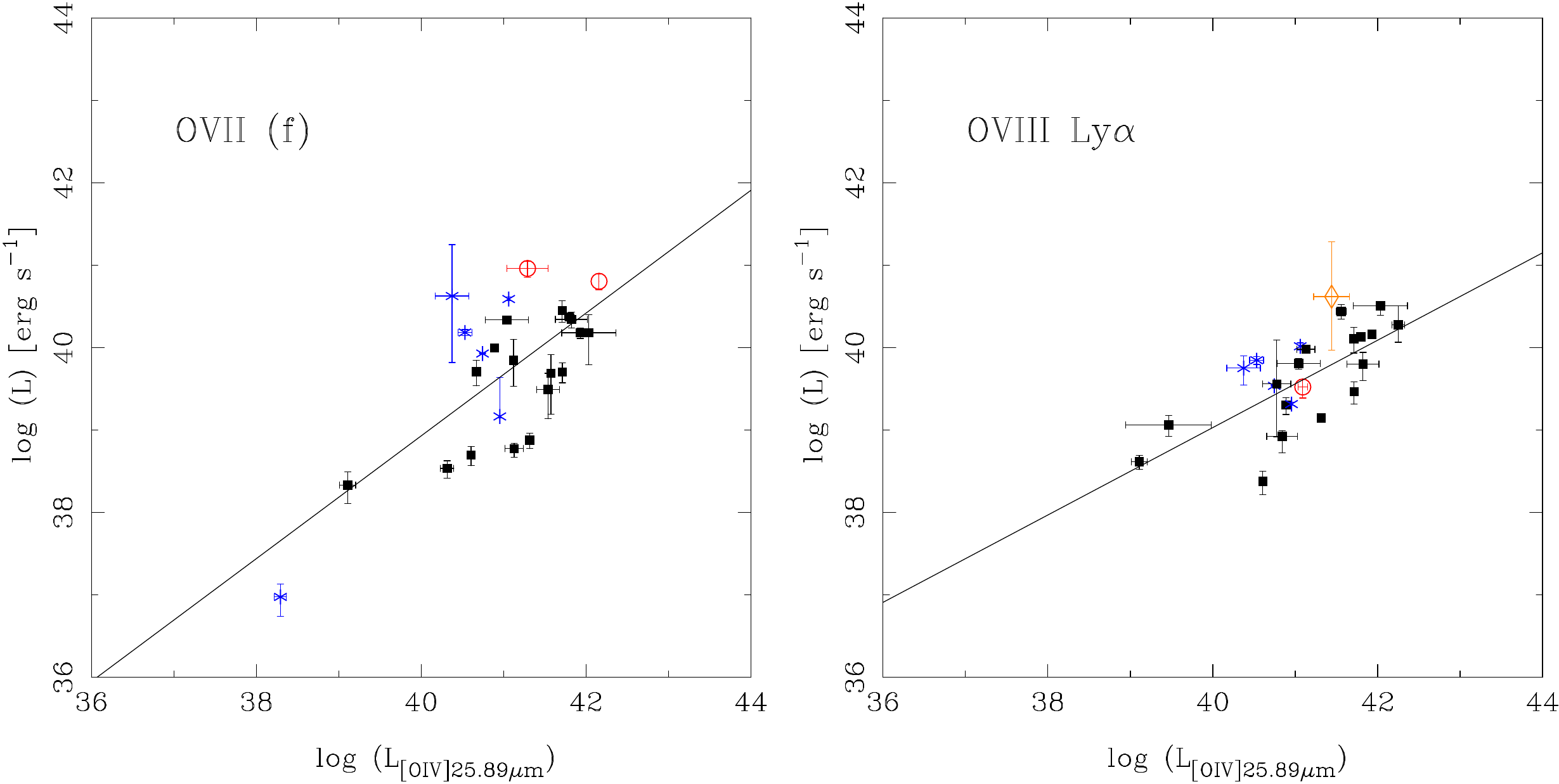}
\includegraphics[angle=270,width=0.8\textwidth]{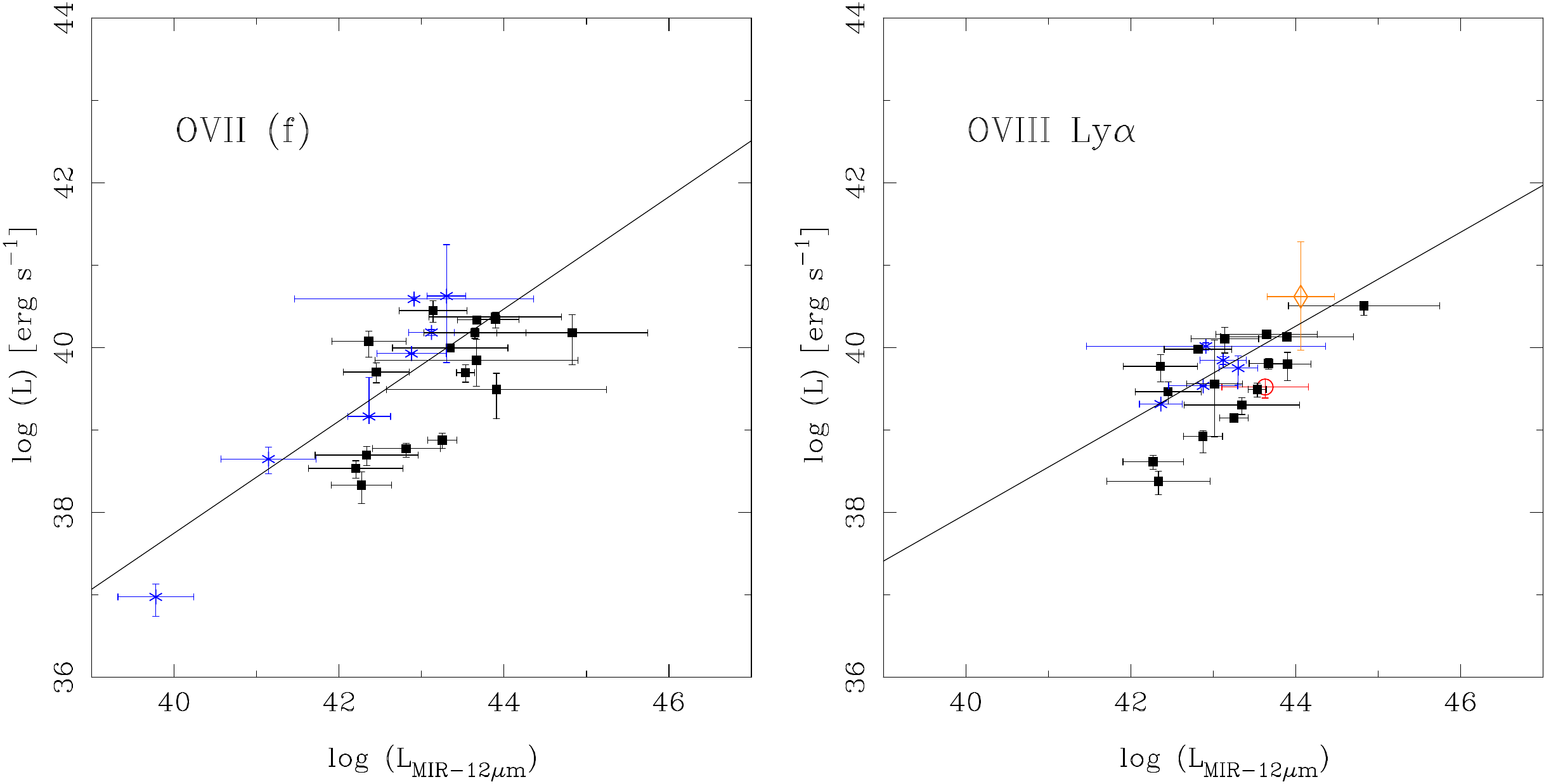}
\caption{X-ray oxygen emission lines vs. indicators of the AGN ionizing power: [OIII]$\lambda5007$ from the optical (top panels), [OIV]25.89\micron~from the IR (middle panels), and MIR 12\micron~continuum (bottom panels). Sources are described in Fig.~\ref{F:LvsC_HX}.}\label{F:proxy-AGN}
\end{figure*}

In Figure~\ref{F:proxy-AGN} we plot the luminosities of OVII~(f) and OVIII~Ly$\alpha$ against the luminosity of three multi-wavelength proxies of the AGN ionizing power: [OIII] in the upper diagrams, [OIV] in the middle pannels and MIR 12\micron~ in the lowest ones. Linear fittings are superimposed on each diagram too; their parameters are listed in Table~\ref{T:correlation}. Each one of the selected proxies holds one statistically significant relationship with the X-ray oxygen lines. These results are in agreement with those of PIE models (Fig.~\ref{F:PIE}): the same central ionizing field heats the gas in the 100-pc scale, and the gas cools down through emission of radiation. In that region ions with very different ionization states co-exist. Under this hypothesis, the places where each emission-line is formed within the NLR are strongly dependent on the gas density profile as a function of distance to the radiation source, according to \citet{bianchi06} and \citet{wang11} and assuming that [OIV] can be formed there in the same way as [OIII] does. 

In this sense, \citet{bianchi19} recently showed that the RPC mechanism \citep{dopita02,draine11}, which naturally drives a density stratification in the gas clouds exposed to an ionizing field, can account for the soft X-ray spectra of CHRESOS's sources. \citet{bianchi19} developed a theoretical differential emission-measure (DEM\footnote{The differential emission-measure distribution is defined as DEM$=d(EM)/d(log\xi)$, where EM is the {\it Emission Measure}: EM$=\int_V n_e^2 dV$, and $V$ is the emitting-gas volume.}) for RPC context which nicely corresponds with the observed DEM in CHRESOS. Since the DEM gathers the contribution of every ionization zone in a nebulae to the total observed flux in a given emission line (and this applies to the entire soft X-ray spectrum), the relationships of Fig.~\ref{F:proxy-AGN} can also be understood in terms of RPC: the development of both the gas density profile and ionization profile within the NLR clouds, may occur in response to the ionizing-field pressure exerted on them.

\subsection{IR SB spectroscopic indicators}\label{SB}

%........FIR 
The very cold dust ($T\lesssim40$~K) emission observed at FIR wavelenghts both in AGNs and SB galaxies is widely accepted as a tracer of SB activity \citep{rodriguez86,rodriguez87,mouri92,hatzi10}. Cold dust is found in molecular clouds surrounding star-forming regions. Obviously, these regions can be located in the vicinity of the central engine ({\it i.e.} circumnuclear regions) and be affected by its radiation field, but even in those cases the heating of such cold gas and dust would be mainly produced by both UV photons and also non-ionizing photons from O-B-recently formed stars \citep{mouri92,magnelli12}, capable of penetrating higher density (lower temperature) gas, and reaching the dusty regions. 
In addition, the presence of high-mass X-ray binaries (HMXB), low-mass X-ray binaries (LMXB), young supernova remnants (SNRs) and cooling hot gas have made the star formation regions great laboratories to study how different emission processes take place. Such an analysis must be necessarily done in a multi-wavelength approach, such as that of \citet{tullmann06}, who showed the relationships among radio (1.4~GHz), FIR (60\micron~and 100\micron), H$\alpha$, B-band, UV and soft X-rays.

While the radiation field of massive stars penetrate the cold phases of the ISM (gas and dust) and heat them triggering the FIR emission, the supernova explosions also accelerates free electrons producing the observed radio syncrotron emission: FIR and radio continuum (1.4~GHz) show a very well known and very well behaved relationship \citep{kruit73,harwit75,helou85,jong85,condon92,tullmann06}.

On the other side of the spectrum, the relationship between the soft X-ray band (0.5-3~keV and 0.3-2~keV) and FIR in SB is also largely known \citep{fabiano89,tullmann06,rosa_gonzalez07}, and even the 2--10~keV has been proposed as an indicator of the star-formation rate in some SB (\citealt{ranalli03}, \citealt[][and references therein]{masHesse08}).

\begin{figure*}
\centering
\includegraphics[angle=270,width=0.8\textwidth]{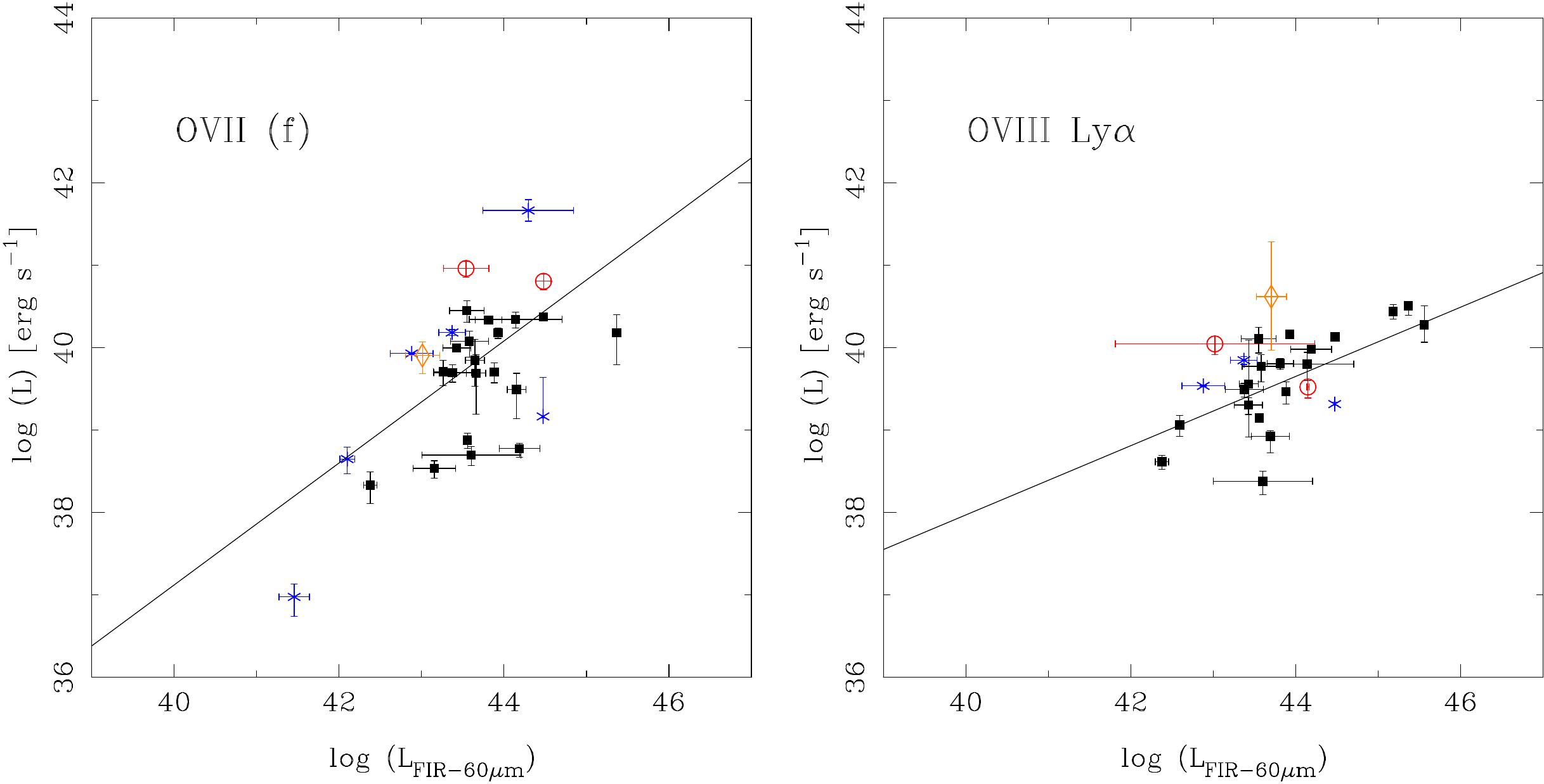}
\includegraphics[angle=270,width=0.8\textwidth]{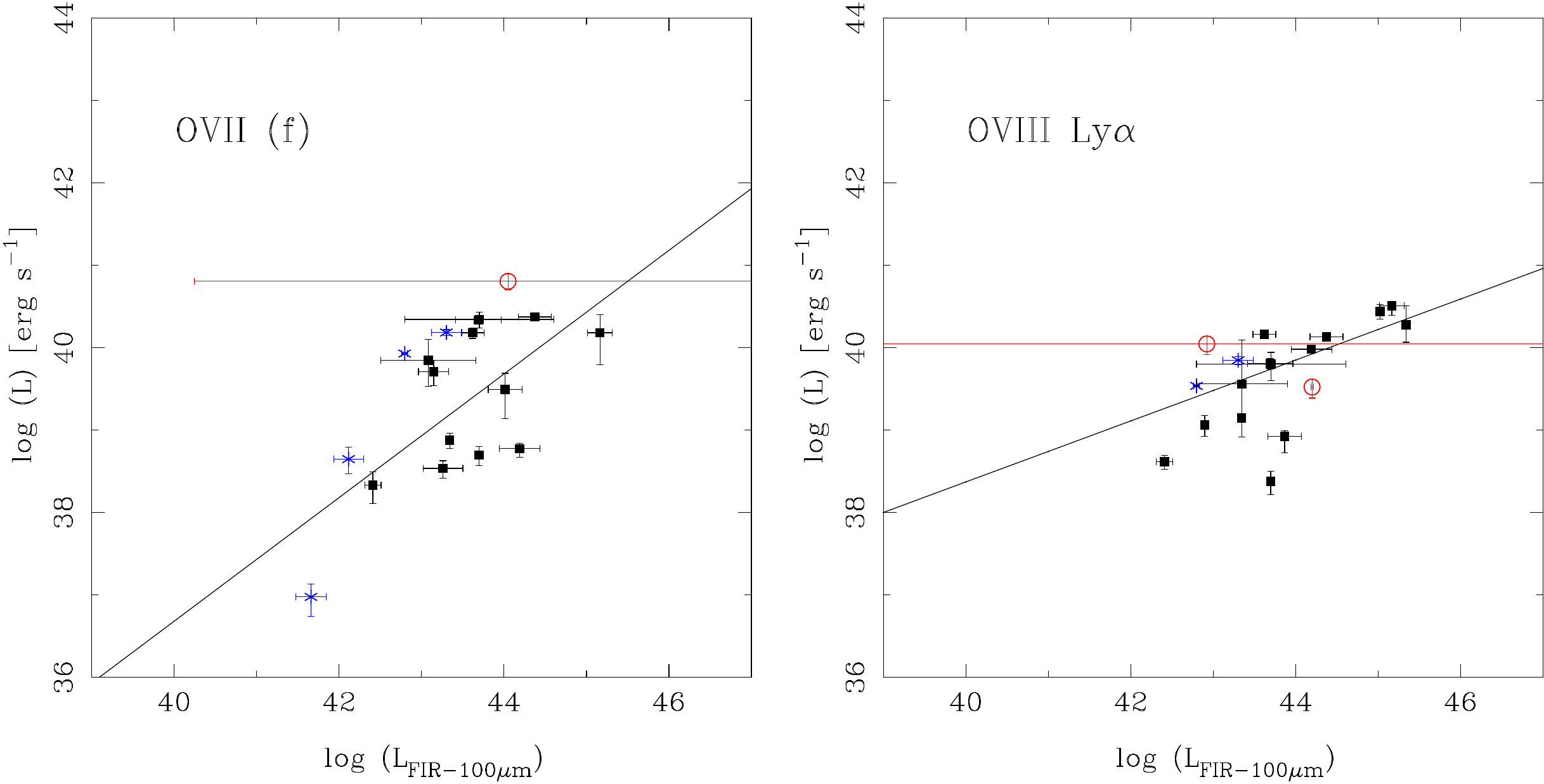}
\caption{X-ray oxygen emission lines vs. proxies of the SB activity: FIR fluxes at 60\micron~ (top panels) and 100\micron~ (bottom panels). Sources are (color-)symbol-coded as in Fig~\ref{F:LvsC_HX}.}\label{F:proxy-SB}
\end{figure*}

As we have already done with proxies of the AGN, we are also interested in probing connections between the X-ray oxygen emission lines and the SB activity throughout their FIR continuum fluxes at 60\micron~and 100\micron. The diagrams are shown in Figure~\ref{F:proxy-SB}. We found the same well-behaved relationship as in the other cases of study. The correlations with 100\micron~luminosity are weaker than those with 60\micron~ luminosity, according to $\rho_S$. The OVIII~Ly$\alpha$ vs. 60\micron~ is the weakest relationship among the statistically significant ones (Table~\ref{T:correlation}). Since our diagrams probe into relationships or connections driven by the SB radiation field, these correlations could be driven by the correlation against the soft X-ray continuum described earlier.

However, the presence of HMXB, LMXB, hot plasma, SNR, and winds accounts for X-ray emission too. It is also known that the X-ray emission in SB can be generated by the injection of mechanical energy into the ISM through SN explosions, shocks and/or stellar winds \citep{cervino02,persic02,masHesse08}. The modelling of \citet{castellanos15} also showed that SN explosions are the main responsible for the injection of mass and energy that produce the observed soft and hard X-ray emission. Then, collisional processes should not be neglected in this analysis. As it was shown in Fig.~\ref{F:CIE}, our model (albeit rather simple) tell us that there is a range of gas temperature where CIE can reproduce the observed OVIII~Ly$\alpha$/OVII~(f) ratios. So, if a contribution from SB activity onto both the formation and observed intensities of OVII~(f) and OVIII~Ly$\alpha$ is indeed relevant, it is likely a product of CIE processes. A stratified nebula would be mandatory in this case, since the critical densities of these lines decrease while the oxygen ionization state increases ($n_{crit}^{[OIII]}=10^{5.845}$~cm$^{-3}$ \citealt{peterson97,rigby09}; $n_{crit}^{[OIV]}=10^{4.06}$~cm$^{-3}$ \citealt{rigby09}; and $n_{crit}^{OVIII}=10^{1.961}$~cm$^{-3}$ \citealt{goddard03}).

%%%%%%%%%%%%%%%%%%%%%%%%%%%%%%%%%%%%%%%%%%%%%%%%%%%%%%%%%%%%%%%%%%%%%%%%%%%%%%%%%%%.
%
%           CONCLUSIONS
%

%%%%%%%%%%%%%%%%%%%%%%%%%%%%%%%%%%%%%%%%%%%%%%%%%%%%%%%%%%%%%%%%%%%%%%%%%%%%%%%%%%%

\section{Summary and Conclusions.}
\label{conclusions}

The most complete soft X-ray oxygen emission-line observations up to now was combined with a variety of continua and emission-line data from the literature in different wavelengths aiming to identify the main process that give raise to such emission. Hard and very hard X-rays, FIR, and MIR continua, as well as optical and IR emission lines were tested against the best-quality measurements of OVII~(f) and OVIII~Ly$\alpha$ so far obtained, through diagnostic diagrams to identify whether their formation is better associated to the AGN or to the SB ionizing power. We have seen that, in the latter case, the ionizing power is almost restricted to collisional ionizing processes since the energy required to photoionize the oxygen ions up to the highest levels cannot be reached by photons from {\it normal} SB radiation fields.

Our theorethical analysis shows that both photoionization and collisional ionization models are capable of predicting the observed OVIII~Ly$\alpha$ to OVII~(f) ratios, for a limited range of parameters, {\it i.e.} ionization parameter in combination with column density, and gas temperature (and ambient density). 

We have combined the X-ray oxygen emission-line luminosities with those of AGN and SB proxies, and with two hard X-ray continuum bands. Regression and correlation analyses were performed in order to find and assess for significant relationships. We have found five (5) statistically significant relationships (at 99\% confidence level) according to their {\it p-values}. Three (3) of them arise with optical/IR proxies of the AGN power in the NLR.

We conclude that the soft X-ray oxygen emission lines (being OVII~(f) and OVIII~Ly$\alpha$ our workhorses) are mainly powered by the AGN, whose radiation field can easily account for the high-energy photons required to let the oxygen 6- or 7-times ionized. A density profile such as $\sim r^{-2}$ \citep{bianchi06,wang11} and the interplay between the ionizing field and the column density $N_H$ \citep{kallman14} play a decisive role in this case. The recent results of \citet{bianchi19} on RPC scenario also point to the AGN as the main source of ionizing photons in the very same CHRESOS sample. Collisional processes principally linked to (but not restricted to) SB scenario, however, cannot be neglected at all and they might represent an important contribution in some cases. Only by enlarging the sample we will be able to both improve our knowledge on the place where these lines are formed, and quantify the role that SB activity plays.

%%%%%%%%%%%%%%%%%%%%%%%%%%%%%%%%%%%%%%%%%%%%%%%%%%%%%%
%% Acknowledgments

\section*{Acknowledgments}

We want to thank to the anonymous referee for helping us to improve the presentation of the paper.V.R. would like to thank to the Committee on Space Research (COSPAR) for having been awarded with the Capacity Building Fellowship Program, and to ESAC/ESA for their support under its visitor programm where this work started. I'm also grateful to Dr. Timothy Kallman for kindly sharing the special version of {\sc xstar} code with us; and to Dr. Daniel Carpintero for his help with the statistics. The present work was supported by grant 11/G153 from the Universidad Nacional de La Plata, PICT-2017-2865 (ANPCyT) and PIP 0102 (CONICET). FG acknowledges support from Athena project number 184.034.002, which is (partly) financed by the Dutch Research Council (NWO). This research made use of the NASA/IPAC Extragalactic Database (NED), which is operated by the Jet Propulsion Laboratory, Caltech, under contract with NASA.

%%%%%%%%%%%%%%%%%%%%%%%%%%%%%%%%%%%%%%%%%%%%%%%%%%%%%%

\section*{Data availability}
The data underlying this article are available in the article and in its online supplementary material.

%%%%%%%%%%%%%%%%%%%%%%%%%%%%%%%%%%%%%%%%%%%%%%%%%%%%%%%%%%%%%%%%%%%%%%%%%%%%%%%%%%%
\bsp

%%%%%%%%%%%%%%%%%%%%%%%%%%%%%%%%%%%%%%%%%%%%%%%%%%%%%%
%% Bibliography

\bibliographystyle{mnras}
\bibliography{alejandria.bib}

%%%%%%%%%%%%%%%%%%%%%%%%%%%%%%%%%%%%%%%%%%%%%%%%%%%%%%

\appendix
\section{Checking for spurious linear fittings: Scrambling Test.}
\label{test}

In order to verify that the luminosity-luminosity correlations in Sect.~\ref{AGN} is not dominated by a distance bias, we have performed the so-called Scrambling Test as proposed by \citet{bianchi09II}, and references therein. The aim of the analysis is as follows: we find a statistically significant relationship between two observed quantities; if we replace one of those quantities by a simulated dataset we expect not to recover the former real, physical, legitimate relationship between the two set of real data because there exist none physical relation between one real data set and the other {\it fake} data set.

We have applied the test onto the best (most significant) relationship we have found, this is OVII~(f) vs. 14--195~keV. Two observed quantities are involved (the luminosities), which depend upon several parameters. For the simulated datasets we keep the parameter whose infuence we are interested in testing for: the distance. To do so, we simulated $10^5$ of these sets, by assigning a ramdomly generated value of OVII(f) {\it Flux} (from a total of $2\times 10^6$) $F^{*}_{OVII(f)}$ to each fixed pair of observed BAT {\it Luminosity} and redshift ($L_{BAT},z$). $F^{*}_{OVII(f)}$ values are then transformed to luminosities ($L^{*}_{OVII(f)}$) (simulated OVII(f) {\it Luminosity}) by using their $z$. The sample size of each new set is the same as that of the real one (21 data-points), otherwise the statistical result would not be comparable.

\begin{figure}
\centering
\includegraphics[angle=0,width=0.4\textwidth]{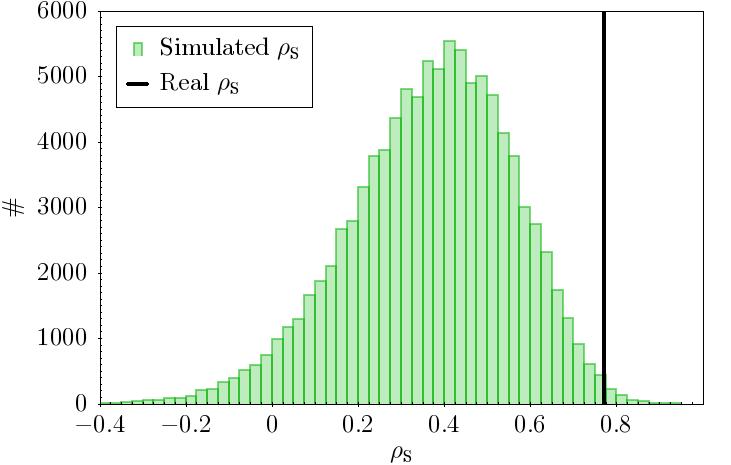}
\includegraphics[angle=0,width=0.4\textwidth]{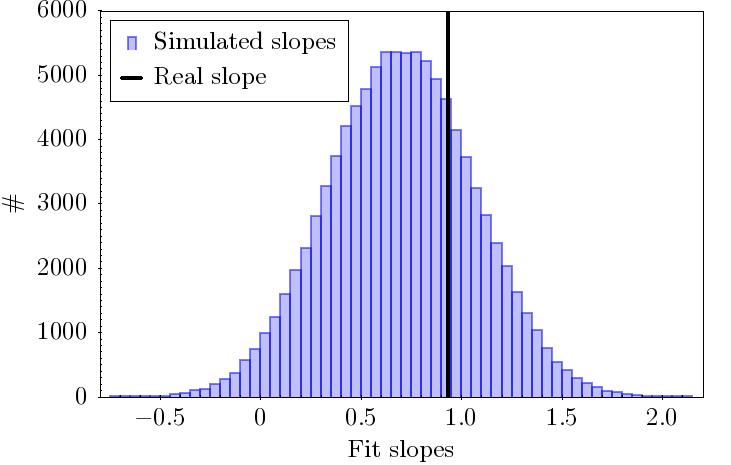}
\caption{Scrambling Test: histogram of $\rho_S$ for the simulated datasets (top) and histogram of fit-slopes (bottom). The vertical lines superimposed on each plot represent $\rho_S$ and the slope for the real dataset (see Table~\ref{T:correlation}).}\label{F:scrambling}
\end{figure}

On each of the $10^5$ simulated datasets we performed the linear fittings and correlations analysis between $L_{BAT}$ and $L^{*}_{OVII(f)}$, that is, the scrambling test. We need to know whether that large amount of correlations are acceptable or not according to the distribution (histogram) of emerging correlation coefficients (Spearman) in comparison with the real, observed one. The results are shown in Figure~\ref{F:scrambling}. The histogram on the top diagram represents the resulting Spearman correlation coefficients ($\rho_S$) of all simulated datasets; a vertical line superimposed on it shows the real one, $\rho^r_S$ (see Table~\ref{T:correlation}). The bottom plot shows the histogram of slopes, for completeness; it shows that the real slope ({\it i.e.} the slope in the observed-data fitting) lies well within the distribution of random slopes. The main result of the test is that the mode of the distribution/histogram of correlation coefficient is $\rho^m_S\sim 0.42$. Since $\rho^m_S < \rho^r_S$ ({\it i.e.} $\rho^m_S$ is less than the real, observed one, 0.77), this means that simulated-data correlations are weak in comparison with the observed one. From the entire sample of $10^5$ simulated datasets, only less than $1000$ have a correlation coefficient higher than $\rho^r_S$. We then obtain a result in agreement with the test's hypothesis: there exist a genuine correlation on observed data which fades away when the simulated data is used. Hence, the probability that the observed $L_{OVII(f)}$ vs $L_{BAT}$ correlation (Fig.\ref{F:LvsC_HX}) would be distance-driven is lower than 1~per~cent.

%*********************************************************************************************************************************************************

\label{lastpage}

\end{document}